\documentclass{amsart}

\newtheorem{theorem}{Theorem}[section]

\theoremstyle{definition}

\theoremstyle{remark}
\newtheorem{remark}[theorem]{Remark}
\numberwithin{equation}{section}
\newcommand{\abs}[1]{\lvert#1\rvert}

\newcommand{\q}{\mathfrak q}
\newcommand{\p}{\mathfrak p}
\newcommand{\Z}{\mathfrak{Z}}

\begin{document}
\title{Multi-Matrix Vector Coherent States}
\author{K. Thirulogasanthar$^{\dagger}$}
\address{$\dagger,\star$~Department of Mathematics and Statistics, Concordia University,
   7141 Sherbrooke Street West, Montreal, Quebec H4B 1R6, Canada }
\email{santhar@cs.concordia.ca}
\author{G. Honnouvo$^{\star}$}
\email{g$\_$honnouvo@yahoo.fr}
\author{A. Krzy\.zak$^{\ddagger}$}
\address{$\ddagger$~Department of Computer Science, Concordia University, 1455 de Maisonneuve Blvd. West, Montreal, Quebec, H3G 1M8, Canada}
\email{krzyzak@cs.concordia.ca}
\date{\today}
\begin{abstract}
A class of vector coherent states is derived with multiple of matrices as vectors in a Hilbert space, where the Hilbert space is taken to be the tensor product of several other Hilbert spaces. As  examples vector coherent states with multiple of quaternions and octonions are given. The resulting generalized oscillator algebra is briefly discussed. Further, vector coherent states for a tensored Hamiltonian system are obtained by the same method. As  particular cases, coherent states are obtained for tensored Jaynes-Cummings type Hamiltonians and for a two-level two-mode generalization of the Jaynes-Cummings model.
\end{abstract}

\maketitle
\pagestyle{myheadings}
\markboth{Multi-Matrix Vector Coherent States}{K. Thirulogasanthar, G. Honnouvo, A. Krzy\.zak}
\section{Introduction}
The definition of the canonical coherent states, CS for short, of
the harmonic oscillator has been generalized in a number of ways
and successfully applied in quantum physics for a long time. For
various generalizations of this concept and their properties one
may consult \cite{AAG, GS, Pr, Gk,NG,KA,AEG}. Hilbert spaces are
the underlying mathematical structure of several physical
phenomena. In general CS form an overcomplete family of vectors in
the separable Hilbert space of the physical problem.

Let $\mathfrak H$ be a separable Hilbert space with an orthonormal basis $\{\phi_{m}\}_{m=0}^{\infty}$ and $\mathbb C$ be the complex plane. For $z\in\mathfrak D$, an open subset of $\mathbb C$, the states
\begin{equation}
\mid z\rangle=\mathcal N(|z|)^{-\frac{1}{2}}\sum_{m=0}^{\infty}\frac{z^{m}}{\sqrt{\rho(m)}}\phi_{m}
\label{eq1}\end{equation}
are said to form a set of CS if
\begin{enumerate}
\item[(a)]
The states $\mid z\rangle$ are normalized,
\item[(b)]
The states $\mid z\rangle$ satisfy a resolution of the identity, that is
\begin{equation}
\int_{\mathfrak D}\mid z\rangle\langle z\mid d\mu=I\label{eq2}
\end{equation}
\end{enumerate}
where $\mathcal N(|z|)$ is the normalization factor, $\{\rho(m)\}_{m=0}^{\infty}$ is a sequence of nonzero positive real numbers, $d\mu$ is an appropriately chosen measure on $\mathfrak{D}$ and $I$ is the identity operator on $\mathfrak H$.

In a recent article \cite{KA}, the labelling parameter $z$ of (\ref{eq1}) was replaced by an $n\times n$ matrix valued function of the form
$Z=A(r)e^{i\zeta\Theta(k)},$ where $A(r)$ and $\Theta(k)$ are $n\times n$ matrices and thereby a class of vector coherent states (VCS for short) was generated as $n$ component vectors in a Hilbert space $\mathfrak{H}=\mathbb C^{n}\otimes\mathfrak{H}_1$, where $\mathfrak{H}_1$ is an abstract separable Hilbert space. In order to get the normalization and resolution of the identity following conditions were imposed on the matrices.
\begin{equation}\label{eq3}
\Theta(k)=\Theta(k)^{\dagger},\;\;\;[A(r),\Theta(k)]=0,\;\;\;[A(r),A(r)^{\dagger}]=0,
\end{equation}
where $\dagger$ stands for the conjugate transpose of a matrix. The square bracket is equal to zero if the matrices commute. A further generalization of the same concept was given in \cite{AEG}, where the VCS were realized as infinite component vectors in a suitable Hilbert space and as special cases VCS were also given as finite component vectors. It is shown in \cite{KA} and \cite{AEG} that the VCS obtained with the complex representation of quaternions are very effective in describing two level atomic systems placed in a single mode electromagnetic field. In particular \cite{AEG} argues that it is a good fit for the Jaynes-Cummings model.\\
Gazeau and Klauder \cite{Gk} proposed a class of temporally stable CS for Hamiltonians with non-degenerate discrete  spectrum by introducing a generalization of the canonical coherent states. At present this class of CS is known as the Gazeau-Klauder CS. Following their method CS were derived for several quantum systems, see for example \cite{AGM,F}. As a further generalization of this concept multidimensional CS were introduced in \cite{NG} and as a special case of this generalization CS were obtained as a tensor product of two canonical CS in the following form
\begin{equation}\label{eq4}
\mid z_{1},z_{2}\rangle=e^{-(|z_{1}|^2+|z_{2}|^2)/2}\sum_{n_1=0}^{\infty}\sum_{n_2=0}^{\infty}\frac{z_{1}^{n_1}}{\sqrt{n_1!}}\frac{z_{2}^{n_{2}}}{\sqrt{n_2!}}\mid n_1,n_2\rangle.
\end{equation}
The result of \cite{NG} can be considered as an extension to the results of \cite{Gk} in a sense that it allows one to construct CS for a system with several degrees of freedom. For example, for a system with two degrees of freedom it can be written as follows:
\begin{eqnarray}\label{add1}
&&\mid J_1,J_2,\gamma_1,\gamma_2\rangle=\mathcal N_1(J_1,J_2)^{-1}\sum_{n_1}\frac{J_{1}^{n_{1}/2}}{\sqrt{\rho_1(n_1)}}e^{-i\gamma_1E_{n_1}}\mathcal N_2(J_2,n_1)^{-1}\\
&&\hspace{2cm}\times\sum_{n_2}\frac{J_{2}^{n_{2}/2}}{\sqrt{\rho_2(n_1,n_2)}}e^{-i\gamma_2E_{n_2}}\mid n_1,n_2\rangle,\nonumber
\end{eqnarray}
where the labelling parameters and the Hilbert space were chosen appropriately for the system and $\rho(n_1),\rho(n_1,n_2), E_{n_1}, E_{n_2}$ are associated with the spectrum ${\sf{E}}_{n_1,n_2}$ of the system. The dependence of the first sum on the other severely restricts one to consider the states of Eq. (\ref{add1}) as a tensor product of two CS. The detailed explanation can be found in \cite{NG}. This procedure is also used in \cite{TN} to obtain CS for a free magnetic Schr\"odinger operator. It should also be mentioned that in \cite{K} Klauder has obtained CS for the hydrogen atom using a similar method of \cite{NG}.\\
Motivated by these recent developments we generalize the results of \cite{KA},\cite{AEG} and \cite{NG} by introducing multi-matrix VCS in the form
\begin{eqnarray}\label{add2}
&&\mid {\bf A}_{1},{\bf A}_{2},...{\bf A}_{\tau},j\rangle=\mathcal N^{-\frac{1}{2}}\sum_{n_1=0}^{\infty}\sum_{n_2=0}^{\infty}...\sum_{n_{\tau}=0}^{\infty}\frac{{\bf A}_{1}^{n_1}}{\sqrt{\rho_1(n_1)}}\\
&&\hspace{3.5cm}\times\frac{{\bf A}_{2}^{n_{2}}}{\sqrt{\rho_2(n_2)}}...\frac{{\bf A}_{\tau}^{n_{\tau}}}{\sqrt{\rho_{\tau}(n_{\tau})}}\chi^{j}\otimes\phi_{n_1}\otimes\phi_{n_2}\otimes...\otimes\phi_{n_{\tau}}\nonumber
\end{eqnarray}
on a Hilbert space $\mathfrak{H}=\mathbb C^{n}\otimes\left[\bigotimes_{k=1}^{\tau}\mathfrak{H}_{k}\right]$, where ${\bf A}_{k}$'s are $n\times n$ matrix valued functions on appropriate domains, $\{\phi_{n_j}\}_{n_j=0}^{\infty}$ is an orthonormal basis of the separable Hilbert space $\mathfrak{H}_j$,~$j=1,2,...\tau$ and $\{\rho_j(n_j)\}_{n_j=0}^{\infty},~j=1,2,...\tau$ are positive sequences of real numbers. A similar result with complex numbers $z_1,z_2,..,z_{\tau}$ may not be hard to obtain because the complex numbers commute with each other. For matrices, the non-commutativity restricts the construction severely. Furthermore, notice that
$$\mid A_{1}(r_1),A_{2}(r_2),...A_{\tau}(r_{\tau}),j\rangle\not=\mid A_{1}(r_1),j\rangle\otimes\mid A_{2}(r_2),j\rangle\otimes...\otimes\mid A_{\tau}(r_\tau),j\rangle.$$
Next we discuss the physical motivation of the generalization (\ref{add2}). It is well known that the most elementary model describing the interaction of two-level atoms with a single quantized field is the Jaynes-Cummings model. This model is exactly solvable in the rotating wave approximation, see \cite{DH,JA, As} and the references listed therein.
Suppose we have a diagonalizable Hamiltonian $H$ for a two-level atom in a single mode cavity field with non-degenerate energies $E_m^k$ and wavefunctions $\psi_m^k$, $k=1,2,~m=0,1,...,\infty$. Let $x_m^k=E_m^k-E_0^k, ~k=1,2,~m=0,1,...\infty$. Let $R(m)=[\text{diag}(x_m^1!,x_m^2!)]^{-\frac{1}{2}}$ and $Z=\text{diag}(z_1,z_2)$ be diagonal matrices, where $x_m^k!=x_1^k...x_m^k$ is the generalized factorial. Assume that the following vectors are normalized and satisfy a resolution of the identity,
\begin{equation}\label{JCS}
\mid Z,k\rangle=\mathcal{N}(Z)^{-\frac{1}{2}}\sum_{m=0}^{\infty}R(m)Z^m \chi_k\otimes\phi_m\in\mathbb{C}^2\otimes\mathfrak{H},~k=1,2,
\end{equation}
where the wavefunctions $\psi_m^k,~k=1,2$ are identified to the basis of $\mathbb{C}^2\otimes\mathfrak{H}$ as $\phi_m^k:=\chi_k\otimes\phi_m,~k=1,2,~m=0,1,2...$
The collection of vectors
$$\mid Z\rangle=\sum_{k=1}^{2}c_k\mid Z,k\rangle\;\;\;\text{with}\;\;\;\abs{c_1}^2+\abs{c_2}^2=1$$
forms a set of CS for the diagonalized Hamiltonian $H_D$. Such an argument for a two-level system leading to the quaternionic VCS of \cite{KA} was given in \cite{AEG}.

 If one neglects losses, multi-mode multi-level generalizations of the Jaynes- Cummings model can be solved exactly using the exact solvability of the Jaynes- Cummings model \cite{JA, Gao}. By assuming the existence of a solvable multi-mode multi-level system, we also construct VCS as CS of a multi-mode multi-level quantum model as an application to the multi-matrix VCS.  In fact, an extension of (\ref{JCS}) to multi-level multi-mode systems requires multi-matrix VCS. For example, let us consider the following model. In the rotating wave approximation the Hamiltonian of a two-level atom in the two-mode field is given by \cite{Gao}
\begin{equation}\label{2J}
H=\sum_{i=1}^{2}\omega_i\left(a_{i}^{\dagger}a_i+\frac{1}{2}\right)+\frac{1}{2}\omega_0\sigma_{3}+g\left(\sqrt{a_1^{\dagger}a_1a_2^{\dagger}a_2}a_1^{\dagger}a_2^{\dagger}\sigma_-+\sigma_+ a_1a_2\sqrt{a_1^{\dagger}a_1a_2^{\dagger}a_2}\right),
\end{equation}
where $\hbar=1$, $a_i,a_i^{\dagger}~(i=1,2)$ are the annihilation and creation operators of the $i$th mode of the field, $\omega_i~(i=1,2)$, $\omega_0=\omega_1+\omega_2$ (i.e, the two-photon resonance process) are the frequencies of the $i$th mode of the radiation field and the atomic transition, $\sigma_3,\sigma_-$ and $\sigma_+$ are the pseudospin operators of the atom and $g$ is the coupling constant. By diagonalizing $H$ one can obtain the spectrum of $H$ (see \cite{Gao} and the references listed there):
\begin{equation}\label{2S}
E_{\pm}^{n,m}=\omega_1(n+1)+\omega_2(m+1)\pm g(n+1)(m+1).
\end{equation}
For the zero coupling ($g=0$) it can be easily seen that if $\frac{\omega_1}{\omega_2}$ is an irrational number then the spectrum is non-degenerate. By appropriately identifying the wave functions of (\ref{2J}) as the basis of separable abstract Hilbert spaces, as an example of the general model, when $g=0$ we shall introduce CS for the Hamiltonian (\ref{2J}) in the following form:
\begin{equation}
\mid Z,\Z,k\rangle=\mathcal{N}(Z,\Z)^{-\frac{1}{2}}\sum_{n=0}^{\infty}R(n)^{-\frac{1}{2}}Z^n\sum_{m=0}^{\infty}R(n,m)^{-\frac{1}{2}}\Z^m\chi_k\otimes\phi_n\otimes\psi_m,~k=1,2,
\end{equation}
where $Z,\Z, R(n), R(n,m)$ are $2\times 2$ diagonal matrices.
In addition to this example, we shall also present VCS for tensored Jaynes-Cummings type Hamiltonians.\\
In order to avoid technicalities, we carry out our general construction with  two $n\times n$ matrices ${\bf A}$ and ${\bf B}$. Later using the real matrix representation of quaternions and octonions we demonstrate it for several matrices. We also generate VCS in the above form with matrix valued $\rho_j(n_j)$'s. Further, we discuss VCS of type (\ref{add1}). We shall briefly discuss generalized oscillator algebras associated to the VCS. Finally, we introduce VCS for multi-level multi-mode quantum systems.
\section{General construction}\label{gen}
Let $\mathfrak {H}_{1}$, $\mathfrak {H}_{2}$ be two separable Hilbert spaces and $\mathbb C$ be the complex plane. Let $\{\phi_{m}\}_{m=0}^{\infty}$,\;$\{\psi_{l}\}_{l=0}^{\infty}$ and $\{\chi^{j}\}_{j=0}^{n}$ be orthonormal bases of the Hilbert spaces in the respective order and of $\mathbb C^{n}$. Denote $\widehat{\mathfrak H}=\mathbb C^{n}\otimes\mathfrak {H}_{1}\otimes\mathfrak {H}_{2},$ where $\otimes$ stands for the tensor product. Then the set of vectors $\{\chi^{j}\otimes\phi_{m}\otimes\psi_{l}\}_{j,m,l}$ serve as an orthonormal basis of $\widehat{\mathfrak H}$.\\
Let $K,K'$ be measure spaces with probability measures $dK$ and $dK'$, respectively, and $R,S$ be a second pair of measure spaces with measures $dR$ and $dS$, respectively. For $(r,k,\zeta)\in  R\times K\times[0,2\pi)$ and  $(s,k',\eta)\in  S\times K'\times[0,2\pi)$, let
\begin{equation}\label{eq5}
{\bf A}=A(r)e^{i\zeta\Theta(k)},\;\;\;\text{and}\;\;\;
{\bf B}=B(s)e^{i\eta\Lambda(k')},
\end{equation}
where $A(r),\Theta(k),B(s),\Lambda(k')$ are measurable $n\times n$ matrix valued functions with the following properties (assumed to hold almost everywhere with respect to the corresponding measure(s)):
\begin{eqnarray}
&&\Theta(k)=\Theta(k)^{\dagger},\;\;\;[A(r),A(r)^{\dagger}]=0,\;\;\;[A(r),\Theta(k)]=0\label{eq6}\\
&&\Lambda(k')=\Lambda(k')^{\dagger},\;\;\;[B(s),B(s)^{\dagger}]=0,\;\;\;[B(s),\Lambda(k')]=0\label{eq7}\\
&&[B(s),A(r)]=0,\;\;\;[\Lambda(k'),A(r)]=0,\;\;\;[B(s),A(r)^{\dagger}]=0,\;\;\;[\Theta(k),B(s)]=0.\label{eq8}
\end{eqnarray}
Let us recall that
\begin{equation}\label{eq9}
[B(s),A(r)]=0,\;[\Lambda(k'),A(r)]=0,\;[B(s),A(r)^{\dagger}]=0,\;\;[\Theta(k),B(s)]=0
\end{equation}
$$\hspace{0.5cm}\Longrightarrow\;[B(s)^{\dagger},A(r)^{\dagger}]=0,\;[\Lambda(k'),A(r)^{\dagger}]=0,\;[B(s)^{\dagger},A(r)]=0,\;\;[\Theta(k),B(s)^{\dagger}]=0
$$

Let $\mathcal D=R\times S\times K\times K'\times[0,2\pi)\times[0,2\pi)$ and define the measure $d\mu(r,s,k,k',\zeta,\eta)=dRdSdKdK'd\zeta d\eta$ on it. For each pair of matrices ${\bf A},{\bf B}$ we define multi-matrix vector coherent states (MVCS) as follows.
\begin{equation}\label{eq10}
\mid {\bf A},{\bf B},j\rangle=\mathcal N(r,s)^{-\frac{1}{2}}\sum_{m=0}^{\infty}\sum_{l=0}^{\infty}\frac{{\bf A}^{m}}{\sqrt{\rho(m)}}\frac{{\bf B}^{l}}{\sqrt{\lambda(l)}}\chi^{j}\otimes\phi_{m}\otimes\psi_{l},
\end{equation}
where the normalization factor $\mathcal N(r,s)$ and the positive
sequences $\rho(m)$,$\lambda(l)$ have to be chosen so that the
MVCS are normalized and give a resolution of identity. In the
sequel we use the following notation: For an $n\times n$ matrix
$A$, we denote $\abs{A}=[AA^{\dagger}]^{\frac{1}{2}}$.\\ In order
to normalize the states (\ref{eq10}) and to obtain a resolution of
the identity we impose the following assumptions on the matrices
${\bf A}$ and ${\bf B}$: Let ${\bf A}$ and ${\bf B}$ be matrices
of the form (\ref{eq5}) satisfying conditions
(\ref{eq6}),(\ref{eq7}) and (\ref{eq8}). Suppose that $\Theta(k)$
and $\Lambda(k')$ have non-zero integer spectrum. With these
conditions the vectors (\ref{eq10}) are normalized to unity as
\begin{equation}\label{unity}
\sum_{j=1}^{n}\langle {\bf A},{\bf B},j\mid {\bf A},{\bf B},j\rangle=1
\end{equation}
with the normalization factor
\begin{equation}\label{3N}
\mathcal
N(r,s)=\sum_{m=0}^{\infty}\sum_{l=0}^{\infty}\frac{\text{Tr}\abs{A(r)^{m}
B(s)^{l}}^2}{\rho(m)\lambda(l)}.
\end{equation}
In order to see this let us start from (\ref{unity})
and establish (\ref{3N}) in the following way:
\begin{eqnarray*}
&&\sum_{j=1}^{n}\langle {\bf A},{\bf B},j\mid {\bf A},{\bf B},j\rangle
=\mathcal N(r,s)^{-1}\sum_{j=1}^{n}\sum_{m=0}^{\infty}\sum_{l=0}^{\infty}\sum_{\nu=0}^{\infty}\sum_{\tau=0}^{\infty}\frac{1}{\sqrt{\rho(m)\rho(\nu)\lambda(l)\lambda(\tau)}}\\
&&\hspace{3cm}\times\langle {\bf A}^{m}{\bf B}^{l}\chi^{j}\otimes\phi_{m}\otimes\psi_{l}\mid {\bf A}^{\nu}{\bf B}^{\tau}\chi^{j}\otimes\phi_{\nu}\otimes\psi_{\tau}\rangle_{\widehat{\mathfrak H}}\\
&=&\mathcal N(r,s)^{-1}\sum_{j=1}^{n}\sum_{m=0}^{\infty}\sum_{l=0}^{\infty}\sum_{\nu=0}^{\infty}\sum_{\tau=0}^{\infty}\frac{1}{\sqrt{\rho(m)\rho(\nu)\lambda(l)\lambda(\tau)}}\\
&&\hspace{3cm}\times\langle {\bf A}^{m}{\bf B}^{l}\chi^{j}\mid {\bf A}^{\nu}{\bf B}^{\tau}\chi^{j}\rangle_{\mathbb C^{n}}\langle \phi_{m}\mid \phi_{\nu}\rangle_{\mathfrak H_{1}}\langle \psi_{l}\mid \psi_{\tau}\rangle_{\mathfrak H_{2}}\\
&=&\mathcal N(r,s)^{-1}\sum_{j=1}^{n}\sum_{m=0}^{\infty}\sum_{l=0}^{\infty}\frac{1}{\rho(m)\lambda(l)}
\langle {\bf A}^{m}{\bf B}^{l}\chi^{j}\mid {\bf A}^{m}{\bf B}^{l}\chi^{j}\rangle_{\mathbb C^{n}}\\
&=&\mathcal N(r,s)^{-1}\sum_{m=0}^{\infty}\sum_{l=0}^{\infty}\frac{\text{Tr}\abs{{\bf A}^{m}{\bf B}^{l}}^2}{\rho(m)\lambda(l)}
=\mathcal N(r,s)^{-1}\sum_{m=0}^{\infty}\sum_{l=0}^{\infty}\frac{\text{Tr}\abs{A(r)^{m} B(s)^{l}}^2}{\rho(m)\lambda(l)},
\end{eqnarray*}
because
\begin{eqnarray*}
&&\langle {\bf A}^{m}{\bf B}^{l}\chi^{j}\mid {\bf A}^{m}{\bf B}^{l}\chi^{j}\rangle_{\mathbb C^{n}}
=\langle {\bf B}^{l\dagger}{\bf A}^{m\dagger}{\bf A}^{m}{\bf B}^{l}\chi^{j}\mid \chi^{j}\rangle_{\mathbb C^{n}}\\
&=&\langle {\bf B}^{l\dagger}A(r)^{m\dagger}A(r)^{m}{\bf B}^{l}\chi^{j}\mid \chi^{j}\rangle_{\mathbb C^{n}}\;\;\;\;[\text{by}\;\; (\ref{eq6})]\\
&=&\langle B(s)^{l\dagger}A(r)^{m\dagger}A(r)^{m}B(s)^{l}\chi^{j}\mid \chi^{j}\rangle_{\mathbb C^{n}}\;\;\;\;[\text{by}\;\; (\ref{eq7})\;\text{and}\;(\ref{eq8})]\\
\end{eqnarray*}
Therefore, we have (\ref{3N}).
For the resolution of the identity, we have
\begin{eqnarray*}
&&\sum_{j=1}^{n}\int_{\mathcal D}\mid {\bf A},{\bf B},j\rangle\langle {\bf A},{\bf B},j\mid d\mu(r,s,k,k',\zeta,\eta)\\
&=&\sum_{j=1}^{n}\sum_{m=0}^{\infty}\sum_{l=0}^{\infty}\sum_{\nu=0}^{\infty}\sum_{\tau=0}^{\infty}\frac{1}{\sqrt{\rho(m)\rho(\nu)\lambda(l)\lambda(\tau)}}\int_{R}\int_{S}\int_{K}\int_{K'}\int_{0}^{2\pi}\int_{0}^{2\pi}\frac{1}{\mathcal N(r,s)}\\
&&\hspace{1cm}\times\mid A(r)^{m}e^{im\zeta\Theta(k)}B(s)^{l}e^{il\eta\Lambda(k')}\chi^{j}\rangle\langle A(r)^{\nu}e^{i\nu\zeta\Theta(k)}B(s)^{\tau}e^{i\tau\eta\Lambda(k')}\chi^{j}\mid\\
&&\hspace{2cm}\otimes\mid\phi_{m}\rangle\langle\phi_{\nu}\mid\otimes\mid\psi_{l}\rangle\langle\psi_{\tau}\mid d\mu(r,s,k,k',\zeta,\eta)\\
\end{eqnarray*}
\begin{eqnarray*}
&=&\sum_{m=0}^{\infty}\sum_{l=0}^{\infty}\sum_{\nu=0}^{\infty}\sum_{\tau=0}^{\infty}\frac{1}{\sqrt{\rho(m)\rho(\nu)\lambda(l)\lambda(\tau)}}\int_{R}\int_{S}\int_{K}\int_{K'}\int_{0}^{2\pi}\int_{0}^{2\pi}\frac{1}{\mathcal N(r,s)}\\
&&\hspace{1cm}\times A(r)^{m}e^{im\zeta\Theta(k)}B(s)^{l}e^{il\eta\Lambda(k')} B(s)^{\tau\dagger}e^{-i\tau\eta\Lambda(k')}A(r)^{\nu\dagger}e^{-i\nu\zeta\Theta(k)}\\
&&\hspace{2cm}\otimes\mid\phi_{m}\rangle\langle\phi_{\nu}\mid\otimes\mid\psi_{l}\rangle\langle\psi_{\tau}\mid d\mu(r,s,k,k',\zeta,\eta)\\
&&\left[\text{because}\;\;\sum_{j=1}^{n}\mid\chi^{j}\rangle\langle\chi^{j}\mid=\mathbb I_{n},\;\;n\times n\;\text{identity matrix}\right]
\end{eqnarray*}
\begin{eqnarray*}
&=&\sum_{m=0}^{\infty}\sum_{l=0}^{\infty}\frac{4\pi^2}{\rho(m)\lambda(l)}\int_{R}\int_{S}\int_{K}\int_{K'}\frac{1}{\mathcal N(r,s)}\\
&&\hspace{1cm}\times A(r)^{m}B(s)^{l} B(s)^{l\dagger}A(r)^{m\dagger}
\otimes\mid\phi_{m}\rangle\langle\phi_{m}\mid\otimes\mid\psi_{l}\rangle\langle\psi_{l}\mid dRdSdKdK'\\
&&\left[\text{by}\;(\ref{eq6}),(\ref{eq7}),(\ref{eq8})\;\;\text{and}\;\;\int_{0}^{2\pi}e^{i(m-\nu)\zeta\Theta(k)}d\zeta=\left\{\begin{array}{ccc}
0&\text{if}&m\not=\nu\\
2\pi&\text{if}&m=\nu
\end{array}\right.\right]\\
&=&\sum_{m=0}^{\infty}\sum_{l=0}^{\infty}\frac{4\pi^2}{\rho(m)\lambda(l)}\int_{R}\int_{S}\frac{\abs{ A(r)}^{2m}\abs{B(s)}^{2l} }{\mathcal N(r,s)}
\otimes\mid\phi_{m}\rangle\langle\phi_{m}\mid\otimes\mid\psi_{l}\rangle\langle\psi_{l}\mid dRdS\\
&&[\text{because}\; dK,\;dK' \;\text{are probability measures, and by (\ref{eq8})}]\\
&=&\mathbb I_{n}\otimes I_{\mathfrak{H}_{1}}\otimes I_{\mathfrak{H}_{2}},
\end{eqnarray*}
provided that the measures $dR$ and $dS$ satisfy
\begin{equation}\label{eq11}
4\pi^2\int_{R}\int_{S}\frac{\abs{ A(r)}^{2m}\abs{B(s)}^{2l} }{\mathcal N(r,s)}dRdS=\rho(m)\lambda(l)\mathbb I_{n}.
\end{equation}
Thereby, the situation can be summarized as follows: The set of
states (\ref{eq10}) forms a set of MVCS if $\mathcal{N}(r,s)$ of
(\ref{3N}) is finite and the condition (\ref{eq11}) is satisfied.
In this case the states are normalized as (\ref{unity}) and a
resolution of the identity is obtained as
$$\sum_{j=1}^{n}\int_{\mathcal D}\mid {\bf A},{\bf B},j\rangle\langle {\bf A},{\bf B},j\mid d\mu(r,s,k,k',\zeta,\eta)=\mathbb I_{n}\otimes I_{\mathfrak{H}_{1}}\otimes I_{\mathfrak{H}_{2}}.$$
In the following let us see some illustrative examples of the main construction.
\subsection{Multi-quaternionic VCS with complex representation}\label{secex}
 As an example of our general construction, in this section we derive MVCS using the complex representation of
quaternions by $2 \times 2$ matrices. Using the basis matrices,
$$\sigma_{0}=\left(\begin{array}{cc}
1&0\\
0&1
\end{array}\right),\;\;
i\sigma_{1}=\left(\begin{array}{cc}
0&i\\
i&0
\end{array}\right),\;\;
-i\sigma_{2}=\left(\begin{array}{cc}
0&-1\\
1&0
\end{array}\right),\;\;
i\sigma_{3}=\left(\begin{array}{cc}
i&0\\
0&-i
\end{array}\right),
$$
where $\sigma_{1},\sigma_{2}$ and $\sigma_{3}$ are the usual Pauli matrices, a general quaternion is written as
$$\q = x_{0}\sigma_{0}+i\underline{x}\cdot\underline\sigma$$
with $x_0 \in \mathbb R , \;\; \underline{x}=(x_{1},x_{2},x_{3})\in \mathbb R^3$ and
$\underline\sigma=(\sigma_{1},-\sigma_{2},\sigma_{3})$. Thus,
\begin{equation}
   \q =\left(\begin{array}{cc}
x_{0}+ix_{3}&-x_{2}+ix_{1}\\
x_{2}+ix_{1}&x_{0}-ix_{3}
\end{array}\right)\; .\label{eq12}
\end{equation}
It is convenient to introduce the polar coordinates:
$$x_{0}=r\cos{\theta},\;\; x_{1}=r\sin{\theta}\sin{\phi}\cos{\psi},\;\;x_{2}=
r\sin{\theta}\sin{\phi}\sin{\psi},\;\;x_{3}=r\sin{\theta}\cos{\phi}\; ,$$
where $r\in [0,\infty),\phi\in [0,\pi] $ and $\theta,\psi\in [0,2\pi)$. In terms of these,
\begin{equation}
\q=A(r)e^{i\theta\sigma(\widehat{n})}
\label{eq13}
\end{equation}
where
\begin{equation}
A(r)=r\mathbb \sigma_0\; , \hspace{.5cm}
 \sigma(\widehat{n})=\left(\begin{array}{cc}
\cos{\phi}&\sin{\phi}e^{i\psi}\\
\sin{\phi}e^{-i\psi}&-\cos{\phi}
\end{array}\right) \hspace{.5cm} \text{and}\hspace{.5cm}\sigma(\widehat{n})^2 = \sigma_0\; .
\label{eq14}
\end{equation}
We denote the field of quaternions by $\mathbb H$. Now let us take two quaternions $\q_{1},\q_{2}$ as follows.
\begin{equation}\label{eq15}
\q_{1}=A(r)e^{i\theta_1\sigma(\widehat{n}_{1})},\;\;\;\q_{2}=B(s)e^{i\theta_2\sigma(\widehat{n}_{2})},
\end{equation}
where
$$A(r)=r\mathbb \sigma_0,\;
 \sigma(\widehat{n}_1)=\left(\begin{array}{cc}
\cos{\phi_1}&\sin{\phi_1}e^{i\psi_1}\\
\sin{\phi_1}e^{-i\psi_1}&-\cos{\phi_1}
\end{array}\right),\;
B(s)=s\mathbb \sigma_0,\;$$
 $$\text{and}\;\;\sigma(\widehat{n}_2)=\left(\begin{array}{cc}
\cos{\phi_2}&\sin{\phi_2}e^{i\psi_2}\\
\sin{\phi_2}e^{-i\psi_2}&-\cos{\phi_2}
\end{array}\right).$$
     The matrices $A(r),B(s),\sigma(\widehat{n}_1)$ and $\sigma(\widehat{n}_2)$ satisfy the conditions (\ref{eq6})-(\ref{eq8}) and the eigenvalues of $\sigma(\widehat{n}_j);~j=1,2$ are $-1$ and $1$.
Let $\{\Phi_{m}\}_{m=0}^\infty$,$\{\Psi_{l}\}_{l=0}^\infty$   be orthonormal bases of  abstract Hilbert spaces
$\mathfrak {H}_{1}$,$\mathfrak {H}_{2}$ respectively, and $\chi^{1},\chi^{2}$
be an orthonormal basis of $\mathbb C^{2}$. Let $\mathcal D=[0,\infty)\times[0,\pi)\times[0,2\pi)\times[0,2\pi)$. Let us take the measure to be
$$d\mu(r,s,\phi_{1},\psi_{1},\theta_{1},\phi_{2},\psi_{2},\theta_{2})=\frac{W(r,s)}{16\pi^2}rdrsds(\sin{\phi_{1}})d\phi_{1}d\psi_{1}d\theta_{1}(\sin{\phi_{2}})d\phi_{2}d\psi_{2}d\theta_{2}$$
on $\mathcal D\times\mathcal D$, where $W(r,s)$ is a density
function to be determined. With the above setup we have a set of
MVCS,
\begin{equation}\label{eq16}
\mid \q_{1},\q_{2},j\rangle=\mathcal N(r,s)^{-\frac{1}{2}}\sum_{m=0}^{\infty}\sum_{l=0}^{\infty}\frac{\q_{1}^{m}}{\sqrt{\rho(m)}}\frac{\q_{2}^{l}}{\sqrt{\lambda(l)}}\chi^{j}\otimes\Phi_{m}\otimes\Psi_{l},\;\;\;j=1,2.
\end{equation}
From (\ref{eq14}),(\ref{eq15}) and (\ref{3N}) the normalization factor can be obtained as
\begin{equation}\label{eq17}
\mathcal N(r,s)=2\sum_{m=0}^{\infty}\sum_{l=0}^{\infty}\frac{r^{2m}s^{2l}}{\rho(m)\lambda(l)}.
\end{equation}
Since
$$\int_{0}^{2\pi}\int_{0}^{2\pi}\int_{0}^{\pi}e^{i(m-\nu)\theta_1\sigma(\widehat{n}_1)}\sin{\phi_1}d\phi_1 d\theta_1 d\psi_1
=\left\{\begin{array}{ccc}
2\pi\;\mathbb I_{2}&{\text{if}}&m=\nu\\
0&{\text{if}}&m\not=\nu
\end{array}\right.$$
and a similar integral holds when the index 1 changes to 2, from (\ref{eq14}),(\ref{eq15}) and (\ref{eq11}),
following the argument of Section \ref{gen}, when $W(r,s)$ satisfies
\begin{equation}\label{eq199}
4\pi^2\int_{0}^{\infty}\int_{0}^{\infty}\frac{W(r,s)r^{2m+1}s^{2l+1} }{\mathcal N(r,s)}drds=\rho(m)\lambda(l).
\end{equation}
we can readily obtain the resolution of the identity
\begin{equation}
 \int_{\mathcal D\times\mathcal D} \mid \q_1,\q_2 , j \rangle\;W(r,s)\;\langle \q_1,\q_2 , j \mid \; d\mu = \mathbb I_{2}\otimes I_{\mathfrak{H}_1}\otimes I_{\mathfrak{H}_2}\; .
\label{eq18}
\end{equation}
In the particular case $\rho(m)=\Gamma(m+1)$ and $\lambda(l)=\Gamma(l+1)$ we have $$\mathcal N(r,s)=2e^{r^2}e^{s^2}=2e^{r^2+s^2}$$
and $W(r,s)=\frac{2}{\pi^2}$ satisfies (\ref{eq199}), which can be seen with the integral representation of the gamma function.
\begin{remark}
$\bullet$~One could solve the moment problem (\ref{eq199}) for various choices of $\rho(m)$ and $\lambda(l)$ using the methods demonstrated in \cite{GPS} and thereby create several classes of multi-quaternionic MVCS.\\
$\bullet$~While keeping conditions (\ref{eq6}),(\ref{eq7}), if we replace conditions (\ref{eq8}) by
\begin{equation}\label{r1}
A(r)A(r)^{\dagger}=A(r)^{\dagger}A(r)=f(r)\mathbb I_{n}\;\;\text{and}\;\;B(s)B(s)^{\dagger}=B(s)^{\dagger}B(s)=g(s)\mathbb I_{n},
\end{equation}
where $f,g$ are functions of $r$ and $s$, we can carry out the construction in a somewhat similar way. We demonstrate it through the following section.
\end{remark}
\subsection{An extension of the quaternion case}
In the case of quaternions we had a more convenient form for the matrix $A(r)$ in which the matrix $A(r)$ is simply $r$ times the identity. Here we will establish MVCS with a little more complicated $A(r)$. This example is, in fact, a generalization of the quaternion case. Let
\begin{eqnarray*}
A(r,s)&=&\left(\begin{array}{cc}
r\mathbb I_{2}&-s\mathbb I_{2}\\
s\mathbb I_{2}&r\mathbb I_{2}
\end{array}\right)
\hspace{.5cm}{\text{and}}\hspace{.5cm}\\
\Theta=\Theta(\widehat{n}_{1},\widehat{n}_{2},\theta)&=&
\left(\begin{array}{cc}
\sigma(\widehat{n}_{1})\sin{\theta} &i\sigma(\widehat{n}_{2})\cos{\theta}\\
-i\sigma(\widehat{n}_{2})\cos{\theta}& \sigma(\widehat{n}_{1})\sin{\theta}
\end{array}\right)
\end{eqnarray*}
where $\widehat{n}_{1}=(n_{11},n_{12},n_{13})$ and $\widehat{n}_{2}=(n_{21},n_{22},n_{23})$ are unit perpendicular vectors and $\sigma(\widehat{n}_{j})= (n_{j1},n_{j2},n_{j3})\cdot (\sigma_{1},\sigma_{2},\sigma_{2})$.  Let
$$\mathcal B_{1}=A(r_{1},s_{1})e^{i\zeta_1\Theta_1}\;\;\text{and}\;\;\mathcal B_{2}=A(r_{2},s_{2})e^{i\zeta_2\Theta_2},$$
where $\Theta_1=\Theta_1(\widehat{n}^{(1)}_{1},\widehat{n}^{(1)}_{2},\theta_1),\;\Theta_2=\Theta_2(\widehat{n}^{(2)}_{1},\widehat{n}^{(2)}_{2},\theta_2)$.
Through straightforward calculations we can see that $A(r_1,s_1),A(r_2,s_2),\Theta_1$ and $\Theta_2$ satisfy conditions (\ref{eq6}),(\ref{eq7}) and (\ref{r1}).
Let $r_1,s_1,r_2,s_2\in [0,\infty)$, $\zeta_1,\theta_1,\zeta_2,\theta_2\in [0,2\pi)$ and $\widehat{n}^{(1)}_{1},\widehat{n}^{(1)}_{2},\widehat{n}^{(2)}_{1},\widehat{n}^{(2)}_{2}\in\Omega$, where $\Omega$ is an open subset of $\mathbb R^3$. For the measure let us take $d\mu=d\nu d\kappa d\Omega$, where $d\nu=r_1s_1r_2s_2dr_1ds_1dr_2ds_2$, $d\kappa=\frac{1}{4\pi^2}d\zeta_1d\zeta_2$  and $d\Omega$ is the probability measure on $\Omega\times\Omega\times[0,2\pi)\times[0,2\pi)$. Let
$$\mathcal D=[\mathbb R^{+}]^{4}\times[0,2\pi)\times[0,2\pi)\times\Omega\times\Omega\times[0,2\pi)\times[0,2\pi).$$
With the above choices and with the usual notations we form a set of MVCS:
\begin{equation}\label{r2}
\mid\mathcal B_{1},\mathcal B_{2},j\rangle=\mathcal N(r_1,s_1,r_2,s_2)^{-\frac{1}{2}}\sum_{m=0}^{\infty}\sum_{l=0}^{\infty}\frac{\mathcal B_{1}^{m}}{\sqrt{\rho(m)}}\frac{\mathcal B_{2}^{l}}{\sqrt{\lambda(l)}}\chi^{j}\otimes\phi_{m}\otimes\psi_{l},\hspace{.5cm}j=1,2,3,4.
\end{equation}
Since
$$\mathcal B_{1}^{m\dagger}\mathcal B_{1}^{m}=(r_{1}^{2}+s_{1}^{2})^{m}\mathbb I_4,\;\;\text{and}\;\;\mathcal B_{2}^{l\dagger}\mathcal B_{2}^{l}=(r_{2}^{2}+s_{2}^{2})^{l}\mathbb I_4,$$
and
$$\langle \mathcal B_{1}^{m}\mathcal B_{2}^{l}\chi^{j}\mid \mathcal B_{1}^{m}\mathcal B_{2}^{l}\chi^{j}\rangle=\langle \mathcal B_{2}^{l\dagger},\mathcal B_{1}^{m\dagger}\mathcal B_{1}^{m}\mathcal B_{2}^{l}\chi^{j}\mid \chi^{j}\rangle=4(r_{1}^{2}+s_{1}^{2})^{m}(r_{2}^{2}+s_{2}^{2})^{l},$$
we have
\begin{equation}\label{r3}
\sum_{j=1}^{4}\langle\mathcal B_{1},\mathcal B_{2},j\mid\mathcal B_{1},\mathcal B_{2},j\rangle=4\mathcal N(r_1,s_1,r_2,s_2)^{-1}\sum_{m=0}^{\infty}\sum_{l=0}^{\infty}\frac{(r_{1}^{2}+s_{1}^{2})^{m}}{\rho(m)}\frac{(r_{2}^{2}+s_{2}^{2})^{l}}{\lambda(l)}.
\end{equation}
Thereby the normalization factor is given by
$$\mathcal{N}(r_1,s_1,r_2,s_2)=4\sum_{m=0}^{\infty}\sum_{l=0}^{\infty}\frac{(r_{1}^{2}+s_{1}^{2})^{m}}{\rho(m)}\frac{(r_{2}^{2}+s_{2}^{2})^{l}}{\lambda(l)}.$$
Further, since
$$\sum_{j=1}^{4}\mid \mathcal B_{1}^{m}\mathcal B_{2}^{l}\chi^{j}\rangle\langle\mathcal B_{1}^{m}\mathcal B_{2}^{l}\chi^{j}\mid= \mathcal B_{1}^{m}\mathcal B_{2}^{l}\left(\sum_{j=1}^{4}\mid \chi^{j}\rangle\langle\chi^{j}\mid\right)(\mathcal B_{1}^{m}\mathcal B_{2}^{l})^{\dagger}=(r_{1}^{2}+s_{1}^{2})^{m}(r_{2}^{2}+s_{2}^{2})^{l}\mathbb I_4,$$
we have
\begin{eqnarray*}
&&\sum_{j=1}^{4}\int_{\mathcal D}W(r_1,s_1,r_2,s_2)\mid \mathcal B_{1},\mathcal B_{2},j\rangle\langle \mathcal B_{1},\mathcal B_{2},j\mid d\mu\\
&=&\sum_{m=0}^{\infty}\sum_{l=0}^{\infty}\frac{1}{\rho(m)\lambda(l)}\int_{0}^{\infty}\int_{0}^{\infty}\int_{0}^{\infty}\int_{0}^{\infty}\frac{W(r_1,s_1,r_2,s_2)(r_{1}^{2}+s_{1}^{2})^{m}(r_{2}^{2}+s_{2}^{2})^{l}}{N(r_1,s_1,r_2,s_2)}d\nu\mathbb I_{4}\\
&&\hspace{4cm}\otimes\mid\phi_m\rangle\langle\phi_{m}\mid\otimes\mid\psi_l\rangle\langle\psi_{l}\mid.
\end{eqnarray*}
Thus a resolution of the identity is satisfied provided that there exists a density $W(r_1,s_1,r_2,s_2)$ satisfying
$$\int_{0}^{\infty}\int_{0}^{\infty}\int_{0}^{\infty}\int_{0}^{\infty}\frac{W(r_1,s_1,r_2,s_2)(r_{1}^{2}+s_{1}^{2})^{m}(r_{2}^{2}+s_{2}^{2})^{l}}{N(r_1,s_1,r_2,s_2)}d\nu=\rho(m)\lambda(l).$$
For the particular case, $\rho(m)=m!$ and $\lambda(l)=l!$ the normalization factor can be easily deduced from (\ref{r3}).
$$\mathcal{N}(r_1,s_1,r_2,s_2)=4e^{(r_{1}^{2}+s_{1}^{2}+r_{2}^{2}+s_{2}^{2})}$$
and a resolution of the identity is obtained with $$W(r_1,s_1,r_2,s_2)=W(r_1,s_1)W(r_2,s_2)$$
where
$$W(r_1,s_1)=\frac{16}{r_{1}^{2}+s_{1}^{2}},\;\;\;W(r_2,s_2)=\frac{16}{r_{2}^{2}+s_{2}^{2}},$$
Which can be seen in the following way:
Since in general
we have
\begin{eqnarray*}
&&\int_{0}^{\infty}\int_{0}^{\infty}\frac{w(r,s)(r^{2}+s^{2})^{m}}{\mathcal N(\mathcal B)}srdsdr\\
&=&\int_{0}^{\infty}\int_{0}^{\infty}\frac{16(r^{2}+s^{2})^{m}}{4(r^{2}+s^{2})e^{r^{2}+s^{2}}}srdsdr\\
&=&\int_{0}^{\infty}\int_{0}^{\infty}\frac{4(r^{2}+s^{2})^{m-1}}{e^{r^{2}}e^{s^{2}}}srdsdr\\
&=&\sum_{j=0}^{m-1}\left(\begin{array}{c}
m-1\\
j
\end{array}\right)
\int_{0}^{\infty}\int_{0}^{\infty}4rs(r^{2})^{m-1-j}(s^{2})^{j}e^{-r^{2}}e^{-s^{2}}drds\\
&=&\sum_{j=0}^{m-1}\left(\begin{array}{c}
m-1\\
j
\end{array}\right)
\int_{0}^{\infty}\left(\int_{0}^{\infty}(r^{2})^{m-1-j}e^{-r^{2}}2rdr\right)(s^{2})^{j}e^{-s^{2}}2sds\\
&=&\sum_{j=0}^{m-1}\left(\begin{array}{c}
m-1\\
j
\end{array}\right)
\Gamma(m-j)\Gamma(j+1)=m!,\;
\end{eqnarray*}
we obtain
$$\sum_{j=1}^{4}\int_{\mathcal D}W(r_1,s_1,r_2,s_2)\mid \mathcal B_{1},\mathcal B_{2},j\rangle\langle \mathcal B_{1},\mathcal B_{2},j\mid d\mu=\mathbb I_{4}\otimes I_{\mathfrak{H}_1}\otimes I_{\mathfrak{H}_1}.$$

\begin{remark}
For $r,s\in[0,\infty)$ and $\theta,\eta\in[0,2\pi)$, if we take our matrices ${\bf A}$ and ${\bf B}$ as,
\begin{equation}\label{ex1}
{\bf A}=e^{i\theta}A(r)\;\;\text{and}\;\;{\bf B}=e^{i\eta}B(s)
\end{equation}
with  conditions,
\begin{equation}\label{ex2}
A(r)A(r)^{\dagger}=A(r)^{\dagger}A(r)=f(r)\mathbb I_{n}\;\;\text{and}\;\;B(s)B(s)^{\dagger}=B(s)^{\dagger}B(s)=g(s)\mathbb I_{n},
\end{equation}
we can carry out the above construction with conditions (\ref{eq6}),(\ref{eq7}) and (\ref{eq8}) replaced by the single condition (\ref{ex2}). The condition (\ref{ex2}) is satisfied by any real Clifford type matrix \cite{key6,AEG}. Following the construction given in Section \ref{gen} one can construct MVCS with any $n\times n$ Clifford matrix. However, since an explicit real matrix representation is known only for quaternions and octonions \cite{key6}, in the following sections we demonstrate it with the real matrix representation of quaternions and octonions.
\end{remark}
\subsection{Multi-quaternionic VCS with real representation}
\label{secq}
Here we present quaternionic MVCS with the real matrix representation of quaternions without imposing any conditions on the matrices. Let
$$\mathbb H=\{\q'=a_{0}+a_{1}i+a_{2}j+a_{3}k\;\mid\;i^{2}=j^{2}=k^{2}=-1,\;ijk=-1,\;a_{0},a_{1},a_{2},a_{3}\in\mathbb R\}$$
be the real quaternion division algebra. It is known that $\mathbb H$ is algebraically isomorphic to the real matrix algebra
$$\mathcal M=\left\{\q'=\left(
\begin{array}{cccc}
a_{0} & -a_{1} & -a_{2} & -a_{3} \\
a_{1} & a_{0} & -a_{3} & a_{2} \\
a_{2} & a_{3} & a_{0} & -a_{1} \\
a_{3} & -a_{2} & a_{1} & a_{0}
\end{array}
\right)\;\;\left|\right.\;\;a_{0},a_{1},a_{2},a_{3}\in\mathbb R\right\}.$$
For a detailed explanation see \cite{key6} and the references therein. For $z=e^{i\theta}\in S^{1}$ and $\q'\in \mathcal M$, let $\q(z)=z\q'$, then we have
$$\q(z)\q(z)^{\dagger}=\q(z)^{\dagger}\q(z)=(a_{0}^{2}+a_{1}^{2}+a_{2}^{2}+a_{3}^{2})\mathbb I_{4}=|\q|^{2}\mathbb I_{4}=|\q'|^{2}\mathbb I_{4}$$
where $|\q'|$ is the norm of the quaternion $\q'$ and $\mathbb I_{4}$ is the $4\times 4$ identity matrix. Let  $\q'(a)$ denotes the matrix representation of $\q'$ with matrix elements $a_{i}$s. Let us consider the Hilbert space $\mathbb C^{4}\otimes\mathfrak {H}_1\otimes\mathfrak {H}_2$.
The quaternion norm can be considered as a continuous function,
$$|\;\cdot\;|:\mathbb H\longrightarrow\mathbb R^{+};\;\;\;\q\mapsto t.$$
Thus, we make the following identification: $\abs{\q_1}=t,\;\abs{\q_2}=s$. Since $z_1=e^{i\theta_1}$ and $z_2=e^{i\theta_2}$, set the measure $d\mu(t,s,\theta_1,\theta_2)=d\theta_1 d\theta_2 tdtsds$ on $\mathcal D=[0,2\pi)\times[0,2\pi)\times\mathbb R^{+}\times\mathbb R^{+}$.
 With the notations of the previous section let us define a set of multi-matrix VCS as follows:\\
For $\;j=1,2,3,4$ the following set of vectors forms a set of MVCS.
\begin{equation}\label{eq20}
\mid \q_1(z_{1},a),\q_2(z_{2},b),j\rangle=\mathcal N(|\q_{1}|,|\q_2|)^{-\frac{1}{2}}\sum_{m=0}^{\infty}\sum_{l=0}^{\infty}\frac{\q_{1}^{m}}{\sqrt{\rho(m)}}\frac{\q_{2}^{l}}{\sqrt{\lambda(l)}}\chi^{j}\otimes\phi_{m}\otimes\phi_{l}.
\end{equation}
Let us find the normalization factor. Since 
\begin{eqnarray*}
&&\sum_{j=1}^{4}\langle \q_1(z_{1},a),\q_2(z_{2},b),j\mid\q_1(z_{1},a),\q_2(z_{2},b),j\rangle\\
&=&\mathcal N(|\q_{1}|,|\q_2|)^{-1}\sum_{j=1}^{4}\sum_{m=0}^{\infty}\sum_{l=0}^{\infty}\frac{1}{\rho(m)\lambda(l)}
\langle \q_{1}^{m}\q_{2}^{l}\chi^{j}\mid \q_{1}^{m}\q_{2}^{l}\chi^{j}\rangle\\
&=&4\mathcal N(|\q_{1}|,|\q_2|)^{-1}\sum_{m=0}^{\infty}\sum_{l=0}^{\infty}\frac{|\q_{1}|^{2m}|\q_2|^{2l}}{\rho(m)\lambda(l)},
\end{eqnarray*}
the normalization factor is given by
\begin{equation}\label{eq21}
\mathcal N(|\q_{1}|,|\q_2|)=4\sum_{m=0}^{\infty}\sum_{l=0}^{\infty}\frac{|\q_{1}|^{2m}|\q_2|^{2l}}{\rho(m)\lambda(l)}.
\end{equation}
Next we obtain a resolution of the identity. Since
 \begin{eqnarray*}
&&\sum_{j=1}^{4}\int_{\mathcal D}\mid \q_1(z_{1},a),\q_2(z_{2},b),j\rangle\langle \q_1(z_{1},a),\q_2(z_{2},b),j\mid d\mu(t,s,\theta_1,\theta_2)\\
&=&\sum_{m=0}^{\infty}\sum_{l=0}^{\infty}\frac{4\pi^2}{\rho(m)\lambda(l)}\int_{0}^{\infty}\int_{0}^{\infty}\frac{W(t,s)t^{2m+1}s^{2l+1}}{\mathcal N(t,s)}dtds \mathbb I_{4}\otimes\mid\phi_{m}\rangle\langle\phi_{m}\mid\otimes\mid\psi_{l}\rangle\langle\psi_{l}\mid\\
&=&\mathbb I_{4}\otimes I_{\mathfrak{H}_1}\otimes I_{\mathfrak{H}_2},
\end{eqnarray*}
provided that there exits a density function $W(t,s)$ such that
\begin{equation}\label{eq22}
4\pi^2\int_{0}^{\infty}\int_{0}^{\infty}\frac{W(t,s)t^{2m+1}s^{2l+1}}{\mathcal N(t,s)}dtds=\rho(m)\lambda(l).
\end{equation}
\begin{remark}
$\bullet$~For the choice $\rho(m)=m!$ and $\lambda(l)=l!$ from (\ref{eq21}) we have $\mathcal N(|\q_{1}|,|\q_2|)=4e^{t^2+s^2}$ and  $W(s,t)=\frac{4}{\pi^2}$  satisfies (\ref{eq22}).\\
$\bullet$~One could easily notice that the whole procedure depends on the following:
\begin{equation}\label{eq23}
\langle\q_{1}^{m}\q_{2}^{l}\chi^{j}\mid\q_{1}^{m}\q_{2}^{l}\chi^{j}\rangle=\langle(\q_{1}^{m}\q_{2}^{l})^{\dagger}\q_{1}^{m}\q_{2}^{l}\chi^{j}\mid\chi^{j}\rangle=\abs{\q_1}^{2m}\abs{\q_2}^{2l}\;\;\text{and}
\end{equation}
\begin{eqnarray}\label{eq24}
\sum_{j=1}^{4}\mid\q_{1}^{m}\q_{2}^{l}\chi^{j}\rangle\langle\q_{1}^{m}\q_{2}^{l}\chi^{j}\mid&=&\q_{1}^{m}\q_{2}^{l}\left(\sum_{j=1}^{4}\mid\chi^{j}\rangle\langle\chi^{j}\mid\right)(\q_{1}^{m}\q_{2}^{l})^{\dagger}\\
&=&\q_{1}^{m}\q_{2}^{l}\q_{2}^{l\dagger}\q_{1}^{m\dagger}=\abs{\q_1}^{2m}\abs{\q_2}^{2l}\mathbb I_{4}.\nonumber
\end{eqnarray}
The identities (\ref{eq23}) and (\ref{eq24}) holds for any number of quaternions, that is, $\q_1\q_2$ can be replaced by $\q_1....\q_{\tau}$, therefore the procedure can be extended to have MVCS on the Hilbert space $\mathbb C^{n}\otimes\left[\bigotimes_{r=1}^{\tau}\mathfrak{H}_{r}\right]$, where $\mathfrak{H}_{r},\;r=1,...,\tau$ are separable Hilbert spaces.
\end{remark}
\subsection{Multi-Octonionic VCS}
Let $\mathbb O$ denote the octonion algebra over the real number field $\mathbb R$. In \cite{key6} it was shown that any $a\in\mathbb O$ has a left matrix representation $\omega(a)$ and a right matrix representation $\nu(a)$ and these representations were given respectively by
$$
\omega(a)=\left(
\begin{array}{cccccccc}
a_{0} & -a_{1} & -a_{2} & -a_{3} & -a_{4} & -a_{5} & -a_{6} & -a_{7} \\
a_{1} & a_{0} & -a_{3} & a_{2} & -a_{5} & a_{4} & a_{7} & -a_{6} \\
a_{2} & a_{3} & a_{0} & -a_{1} & -a_{6} & -a_{7} & a_{4} & a_{5} \\
a_{3} & -a_{2} & a_{1} & a_{0} & -a_{7} & a_{6} & -a_{5} & a_{4} \\
a_{4} & a_{5} & a_{6} & a_{7} & a_{0} & -a_{1} & -a_{2} & -a_{3} \\
a_{5} & -a_{4} & a_{7} & -a_{6} & a_{1} & a_{0} & a_{3} & -a_{2} \\
a_{6} & -a_{7} & -a_{4} & a_{5} & a_{2} & -a_{3} & a_{0} & a_{1} \\
a_{7} & a_{6} & -a_{5} & -a_{4} & a_{3} & a_{2} & -a_{1} & a_{0}
\end{array}
\right)$$
and
$$\nu(a)=\left(
\begin{array}{cccccccc}
a_{0} & -a_{1} & -a_{2} & -a_{3} & -a_{4} & -a_{5} & -a_{6} & -a_{7} \\
a_{1} & a_{0} & a_{3} & -a_{2} & a_{5} & -a_{4} & -a_{7} & a_{6} \\
a_{2} & -a_{3} & a_{0} & a_{1} & a_{6} & a_{7} & -a_{4} & -a_{5} \\
a_{3} & a_{2} & -a_{1} & a_{0} & a_{7} & -a_{6} & a_{5} & -a_{4} \\
a_{4} & -a_{5} & -a_{6} & -a_{7} & a_{0} & a_{1} & a_{2} & a_{3} \\
a_{5} & a_{4} & -a_{7} & a_{6} & -a_{1} & a_{0} & -a_{3} & a_{2} \\
a_{6} & a_{7} & a_{4} & -a_{5} & -a_{2} & a_{3} & a_{0} & -a_{1} \\
a_{7} & -a_{6} & a_{5} & a_{4} & -a_{3} & -a_{2} & a_{1} & a_{0}
\end{array}
\right). $$
Let $z=e^{i\theta}\in S^{1}$ and
$$\omega(a,z)=z\omega(a)\hspace{0.5cm}\text{and}\hspace{0.5cm}\nu(a,z)=z\nu(a).$$
Then
\begin{eqnarray*}
\omega(a,z)\omega(a,z)^{\dagger}&=&\omega(a,z)^{\dagger}\omega(a,z)=\nu(a,z)\nu(a,z)^{\dagger}=\nu(a,z)^{\dagger}\nu(a,z)\\
&=&a_{0}^{2}+a_{1}^{2}+a_{2}^{2}+a_{3}^{2}+a_{4}^{2}+a_{5}^{2}+a_{6}^{2}+a_{7}^{2}=\|a\|^{2}\mathbb I_{8}.
\end{eqnarray*}
Thus, identities similar to (\ref{eq23}) and (\ref{eq24}) can be obtained for $\omega(a,z)$ and $\nu(a,z)$. Now, with  previous notations we can have MVCS as
\begin{eqnarray*}
&&\mid\omega(a_1,z_1),\omega(a_2,z_2),...,\omega(a_{\tau},z_{\tau}),j\rangle,\\
&&\mid\nu(a_1,z_1),\nu(a_2,z_2),...,\nu(a_{\tau},z_{\tau}),j\rangle,\;\;\;\text{or}\\
&&\mid\nu(a_1,z_1),\omega(a_2,z_2),\nu(a_3,z_3),\omega(a_4,z_4)...,\nu(a_{\tau},z_{\tau}),j\rangle.
\end{eqnarray*}
In the last set of states we can mix $\omega(a,z)$ and $\nu(a,z)$ in any order.
\section{Multi-matrix VCS with matrix $\rho(m)$}
Here we consider a class of MVCS in the following form
\begin{equation}\label{rh1}
\mid Z_{1},Z_{2},..,Z_{\tau},j\rangle=\mathcal N^{-\frac{1}{2}}\sum_{n_1=0}^{\infty}...\sum_{n_{\tau}=0}^{\infty}R_{1}(n_1)Z_{1}^{n_1}...R_{\tau}(n_{\tau})Z_{\tau}^{n_{\tau}}\chi^{j}\otimes\phi_{n_1}\otimes...\otimes\phi_{n_{\tau}},
\end{equation}
where $R_{k}(n_{k})$ and $Z_{k}$ are $n\times n$ matrices for all $k=1,...,\tau$. In order to get a normalization and a resolution of identity, we need to take the $Z_{k}$s in a certain way and we have to impose conditions on all the matrices. For example, for $k=1,...,\tau$ let $C_k$ be a real $n\times n$ Clifford matrix, $Z_k=z_kC_k$ with $z_k\in S^{1}$ and $R_k(n_k)$ be a $n\times n$ matrix satisfying
\begin{equation}\label{Rk}
R_{k}(n_{k})R_{k}(n_{k})^{\dagger}=R_{k}(n_{k})^{\dagger}R_{k}(n_{k})=f_{k}(n_{k})\mathbb I_{n},
\end{equation}
where $f_k(n_k)$ is a real valued function of $n_k$. Then following the argument of Section  \ref{gen} one can construct MVCS. For simplicity, we work it out in detail with the real matrix representation of quaternions. Let us take $\q_{k}$ to be as in Section \ref{secq} for all $k=1,2,..,\tau$ and
\begin{equation}\label{ZZ}
Z_{k}=\q_{k}(\theta_{k})=\q_{k}e^{i\theta_{k}}.
\end{equation}
Further, let us take $R_{k}(n_{k})$ such that
\begin{equation}\label{RRR}
R_{k}(n_{k})R_{k}(n_{k})^{\dagger}=R_{k}(n_{k})^{\dagger}R_{k}(n_{k})=f_{k}(n_{k})\mathbb I_{4}=\frac{1}{n_{k}!}\mathbb I_{4},\;\;\;k=1,2,...,\tau.
\end{equation}
For example one such possible $R_{k}(n_{k})$ is
$$R_{k}(n_{k})=\frac{1}{\sqrt{n_{k}!}}\left(\begin{array}{cc}
\cos{(x_{k})}\mathbb I_{2}&-\sin{(x_{k})}\mathbb I_{2}\\
\sin{(x_{k})}\mathbb I_{2}&\cos{(x_{k})}\mathbb I_{2}\end{array}\right),$$
where $x_1,...,x_{\tau}$ are fixed.\\
For $k=1,...,\tau$ let $Z_k$ be as in (\ref{ZZ}) and $R_k(n_k)$ be as in (\ref{RRR}) then the vectors (\ref{rh1}) form a set of MVCS.
Since
\begin{eqnarray*}
&&\langle R_{1}(n_{1})\q_{1}(\theta_{1})^{n_1}...R_{\tau}(n_{\tau})\q_{\tau}(\theta_{\tau})^{n_{\tau}}\chi^{j}\mid R_{1}(n_{1})\q_{1}(\theta_{1})^{n_1}...R_{\tau}(n_{\tau})\q_{\tau}(\theta_{\tau})^{n_{\tau}}\chi^{j}\rangle\\
&=&\langle \q_{\tau}(\theta_{\tau})^{n_{\tau}\dagger}R_{\tau}(n_{\tau})^{\dagger}...\q_{1}(\theta_{1})^{n_1\dagger}R_{1}(n_{1})^{\dagger}R_{1}(n_{1})\q_{1}(\theta_{1})^{n_1}...R_{\tau}(n_{\tau})\q_{\tau}(\theta_{\tau})^{n_{\tau}}\chi^{j}\mid\chi^{j} \rangle\\
&=&\frac{\abs{\q_{1}}^{2n_1}...\abs{\q_{\tau}}^{2n_{\tau}}}{n_1!...n_{\tau}!}\langle\chi^{j}\mid\chi^{j}\rangle
\end{eqnarray*}
and
\begin{eqnarray*}
&&\sum_{j=1}^{4}\mid R_{1}(n_{1})\q_{1}(\theta_{1})^{n_1}...R_{\tau}(n_{\tau})\q_{\tau}(\theta_{\tau})^{n_{\tau}}\chi^{j}\rangle\langle R_{1}(l_{1})\q_{1}(\theta_{1})^{l_1}...R_{\tau}(l_{\tau})\q_{\tau}(\theta_{\tau})^{l_{\tau}}\chi^{j}\mid\\
&=&e^{i(n_1-l_1)\theta_1}...e^{i(n_{\tau}-l_{\tau})\theta_{\tau}}R_{1}(n_{1})\q_{1}^{n_1}...R_{\tau}(n_{\tau})\q_{\tau}^{n_{\tau}}\left(\sum_{j=1}^{n}\mid\chi^{j}\rangle\langle\chi^{j}\mid\right)\\
&&\hspace{3cm}\times \q_{\tau}^{l_{\tau}\dagger}R_{\tau}(l_{\tau})^{\dagger}...\q_{1}^{l_1\dagger}R_{1}(l_1)^{\dagger}\\
&=&e^{i(n_1-l_1)\theta_1}...e^{i(n_{\tau}-l_{\tau})\theta_{\tau}}R_{1}(n_{1})\q_{1}^{n_1}...R_{\tau}(n_{\tau})\q_{\tau}^{n_{\tau}}
\times \q_{\tau}^{l_{\tau}\dagger}R_{\tau}(l_{\tau})^{\dagger}...\q_{1}^{l_1\dagger}R_{1}(l_1)^{\dagger}
\end{eqnarray*}
we have the normalization factor
\begin{equation}\label{n-1}
\mathcal N=\sum_{j=1}^{4}\sum_{n_1=0}^{\infty}...\sum_{n_{\tau}=0}^{\infty}\frac{\abs{\q_{1}}^{2n_1}...\abs{\q_{\tau}}^{2n_{\tau}}}{n_1!...n_{\tau}!}\langle\chi^{j}\mid\chi^{j}\rangle=4e^{\abs{\q_{1}}^2+...+\abs{\q_{\tau}}^2}.
\end{equation}
Set $d\mu=\frac{4}{(\pi)^{\tau}}\abs{\q_1}...\abs{\q_{\tau}}d\abs{\q_1}...d\abs{\q_{\tau}}d\theta_1...d\theta_{\tau}.$ With this measure a resolution of the identity is obtained as follows:
\begin{eqnarray*}
&&\sum_{j=1}^{4}\int_{0}^{\infty}...\int_{0}^{\infty}\int_{0}^{2\pi}...\int_{0}^{2\pi}
\mid \q_{1}(\theta_{1}),...,\q_{\tau}(\theta_{\tau}),j\rangle\langle \q_{1}(\theta_{1}),...,\q_{\tau}(\theta_{\tau})j\mid d\mu\\
&=&\sum_{n_1=0}^{\infty}...\sum_{n_{\tau}=0}^{\infty}\sum_{l_1=0}^{\infty}...\sum_{l_{\tau}=0}^{\infty}\int_{0}^{\infty}...\int_{0}^{\infty}\int_{0}^{2\pi}...\int_{0}^{2\pi}\frac{1}{\mathcal N}\\
&&\;\times e^{i(n_1-l_1)\theta_1}...e^{i(n_{\tau}-l_{\tau})\theta_{\tau}}R_{1}(n_{1})\q_{1}^{n_1}...R_{\tau}(n_{\tau})\q_{\tau}^{n_{\tau}}\chi^{j}
\times \q_{\tau}^{l_{\tau}\dagger}R_{\tau}(l_{\tau})^{\dagger}...\q_{1}^{l_1\dagger}R_{1}(l_1)^{\dagger}d\mu\\
&&\hspace{3cm}\otimes\mid\phi_1\rangle\langle\phi_{1}\mid\otimes...\otimes\mid\phi_{\tau}\rangle\langle\phi_{\tau}\mid\\
&=&\sum_{n_1=0}^{\infty}...\sum_{n_{\tau}=0}^{\infty}2^{\tau}\int_{0}^{\infty}...\int_{0}^{\infty}\frac{\abs{\q_{1}}^{2n_1}...\abs{\q_{\tau}}^{2n_{\tau}}}{n_1!...n_{\tau}!}e^{-\abs{\q_{1}}^2}...e^{-\abs{\q_{\tau}}^2}\abs{\q_1}...\abs{\q_{\tau}}d\abs{\q_1}...d\abs{\q_{\tau}}\\
&&\otimes\mid\phi_1\rangle\langle\phi_{1}\mid\otimes...\otimes\mid\phi_{\tau}\rangle\langle\phi_{\tau}\mid\\
&=&\mathbb I_{4}\otimes I_{\mathfrak{H}_1}\otimes...\otimes I_{\mathfrak{H}_{\tau}}.
\end{eqnarray*}
\begin{remark}
Usually the $\rho(m)$ of (\ref{eq1}) is a positive sequence of real numbers and in getting a resolution of the identity we end up at solving a moment problem with positive moments. When we replace the $\rho(m)$ by a matrix, one could expect to replace it by a positive definite matrix. The matrices we have used are not positive definite. But one can observe that, at the end of the calculations, we end up with a moment problem with positive moments. In this sense, it really does not matter with which matrix we start with but what does matter is the final moment problem.
\end{remark}
\section{ Summations depending on one another}\label{dep}
So far we have demonstrated several classes of MVCS as a generalization of (\ref{eq4}). In this section we present MVCS as a generalization to the states (\ref{add1}), where the summations depend one another. First we give a general construction and then we discuss an example using the complex representation of quaternions.\\
Let ${\bf A}, {\bf B}$ be matrices as in Section \ref{gen} with all the assumptions of Section \ref{gen} on them. Then the following set of vectors forms a set of MVCS.
\begin{equation}\label{sum1}
\mid {\bf A},{\bf B},j\rangle=\mathcal N_{1}(r,s)^{-\frac{1}{2}}\sum_{m=0}^{\infty}\frac{{\bf A}^{f(m)}}{\sqrt{\rho_1(m)}} \mathcal N_{2}(s,m)^{-\frac{1}{2}}\sum_{l=0}^{\infty}\frac{{\bf B}^{g(l)}}{\sqrt{\rho_2(m,l)}}\chi^{j}\otimes\phi_m\otimes\psi_{l},
\end{equation}
where $f,g$ are some functions of $m$ and $l$ respectively. Following the steps of Section \ref{gen} we can get the normalization factor
\begin{equation}\label{sum2}
\mathcal N_{1}(r,s)=\sum_{m=0}^{\infty}\sum_{l=0}^{\infty}\frac{\text{Tr}\abs{A(r)^{f(m)}B(s)^{g(l)}}^2}{\mathcal N_{2}(s,m)\rho_1(m)\rho_2(m,l)}.
\end{equation}
Again by following the steps of Section \ref{gen} we can have the resolution of the identity
\begin{eqnarray*}
&&\sum_{j=1}^{n}\int_{\mathcal D}\mid {\bf A},{\bf B},j\rangle\langle {\bf A},{\bf B},j\mid d\mu(r,s,k,k',\zeta,\eta)\\
&&=\sum_{m=0}^{\infty}\sum_{l=0}^{\infty}\frac{4\pi^2}{\rho_1(m)\rho_2(m,l)}\int_{R}\int_{S}\frac{\abs{{\bf A}}^{2f(m)}\abs{{\bf B}}^{2g(l)}}{\mathcal N_1(r,s)\mathcal N_{2}(s,m)} \otimes\mid\phi_{m}\rangle\langle\phi_{m}\mid\otimes\mid\psi_{l}\rangle\langle\psi_{l}\mid \\
&&\hspace{6cm}\times\lambda_1(r)\lambda_2(r,s) dRdS\\
&&=\mathbb I_{n}\otimes I_{\mathfrak{H}_1}\otimes I_{\mathfrak{H}_2}
\end{eqnarray*}
provided that there are densities $\lambda_1(r)$ and $\lambda_2(s,m)$ such that
\begin{equation}\label{sum3}
4\pi^2\int_{R}\int_{S}\frac{\abs{{\bf A}}^{2f(m)}\abs{{\bf B}}^{2g(l)}}{\mathcal N_1(r,s)\mathcal N_{2}(s,m)}\lambda_1(r)\lambda_2(s,m)dRdS=\rho_1(m)\rho_2(m,l)\mathbb I_{n}.
\end{equation}
where $R,S$ and $dR,dS$ are as in Section \ref{gen}.\\
Now let us see an illustrative example with the complex representation of quaternions.
Take the quaternions as in (\ref{eq13}). Let
$$\mathfrak{G}_1=\{\q=A(r)e^{i\zeta\sigma(\widehat{n_1})}\;:\;r\in[0,\infty)\},\;\;\;\mathfrak{G}_2=\{\p=A(s)e^{i\eta\sigma(\widehat{n_2})}\;:\;s\in[0,1)\}.$$
The variable $s$ is restricted on the set $[0,1)$, the rest of the variables obey the conditions of Section \ref{secex}. For $\q\in\mathfrak{G}_1$, $\p\in\mathfrak{G}_2$, $f(m)=\frac{m}{2},g(l)=\frac{l}{2}$ with
$${\bf A}=\q,\;\;{\bf B}=\p,\;\;\rho_2(m,l)=\left(\begin{array}{c}m+l\\l\end{array}\right)^{-1},\;\;\rho_1(m)=m!\;\;\text{and}\;\;j=1,2$$
 let us consider the states (\ref{sum1}). From (\ref{sum2}) and the properties of quaternions we obtain
\begin{equation}\label{sum4}
\mathcal N_{1}(r,s)=2\sum_{m=0}^{\infty}\sum_{l=0}^{\infty}\left(\begin{array}{c}m+l\\l\end{array}\right)\frac{\mathcal N_{2}(s,m)^{-1}r^{m}s^{m}}{m!}
\end{equation}
where
\begin{equation}\label{sum5}
N_{2}(s,m)=\sum_{l=0}^{\infty}\left(\begin{array}{c}m+l\\l\end{array}\right)s^{m}=(1-s)^{-(m+1)}.
\end{equation}
Therefore we get
\begin{equation}\label{sum6}
\mathcal N_{1}(r,s)=2\sum_{m=0}^{\infty}\frac{(1-s)^{(m+1)}r^{m}}{m!}=(1-s)e^{r(1-s)}
\end{equation}
For a resolution of the identity (\ref{sum3}) reduces to
\begin{equation}\label{sum7}
4\pi^2\int_{0}^{\infty}\int_{0}^{1}\frac{r^ms^l(1-s)^{m}}{e^{r(1-s)}}\lambda_1(r)\lambda_2(r,s,m)drds=m!\left(\begin{array}{c}m+l\\l\end{array}\right)^{-1}.
\end{equation}
Since
\begin{equation}\label{sum8}
m\int_{0}^{1}s^{l}(1-s)^{m-1}ds=\left(\begin{array}{c}m+l\\l\end{array}\right)^{-1}
\end{equation}
we take
$$\lambda_2(r,s,m)=\frac{m}{2\pi(1-s)}e^{rs}.$$
Thereby from (\ref{sum7}) we only have to solve
\begin{equation}\label{sum9}
2\pi\int_{0}^{\infty}r^{m}e^{-r}\lambda_1(r)dr=m!.
\end{equation}
By the definition of the Gamma function, we obtain (\ref{sum9}) when $\lambda_1(r)=\frac{1}{2\pi}$. Thus we have a resolution of the identity.
\section{Generalized oscillator algebra}
For the states (\ref{eq1}), a general way of defining an associated oscillator algebra is as follows \cite{AAG}: Let
$$x_{m}=\frac{\rho(m)}{\rho(m-1)},\;\;\;\forall m\geq 1,\;\;\;\text{and}\;\;\;x_{0}!=1$$
then $\rho(m)=x_m!$. A set of operators, namely annihilation, creation and number operators on the basis vectors $\{\phi_m\}$ is defined as
$$a\phi_m=\sqrt{x_m}\phi_{m-1},\;\;\; a^{\dagger}\phi_m=\sqrt{x_{m+1}}\phi_{m+1},\;\;\;N\phi_m=x_m\phi_{m}.$$
The states (\ref{eq1}) satisfy the relation $a\mid z\rangle=z\mid z\rangle$, that is, the CS are eigenvectors of the operator $a$. These three operators together with the identity operator, under the commutator bracket, generate a Lie algebra, which is the so-called generalized oscillator algebra. In analogy with the above case, for the states (\ref{add2}) let us take
$$x_{n_j}=\frac{\rho_j(n_j)}{\rho_j(n_j-1)},\;\;\;x_0!=1;\;\;\;\forall j=1,...,\tau $$
Let us define a similar set of operators for the basis vectors $\{\chi^{j}\otimes\phi_{n_1}\otimes...\otimes\phi_{n_{\tau}}\}$ as follows:
\begin{eqnarray}\label{ooo}
&&A \chi^{j}\otimes\phi_{n_1}\otimes...\otimes\phi_{n_{\tau}}=\sqrt{x_{n_1}x_{n_2}...x_{n_{\tau}}}\chi^{j}\otimes\phi_{n_1-1}\otimes...\otimes\phi_{n_{\tau}-1}\\
&&A^{\dagger} \chi^{j}\otimes\phi_{n_1}\otimes...\otimes\phi_{n_{\tau}}=\sqrt{x_{n_1+1}x_{n_2+1}...x_{n_{\tau}+1}}\chi^{j}\otimes\phi_{n_1+1}\otimes...\otimes\phi_{n_{\tau}+1}\label{mmb}\nonumber\\
&&N\chi^{j}\otimes\phi_{n_1}\otimes...\otimes\phi_{n_{\tau}}=x_{n_1}x_{n_2}...x_{n_{\tau}}\chi^{j}\otimes\phi_{n_1}\otimes...\otimes\phi_{n_{\tau}}.\nonumber
\end{eqnarray}
If
\begin{equation}\label{mb}
x_{n_j}-x_{n_j-1}=c,\;\text{a constant}\;\;\;\forall j=1,...,\tau,\;\forall n_j
\end{equation}
we get
$$[A,A^{\dagger}]=cI,\;\;\;[N,A]=-cA,\;\;[N,A^{\dagger}]=cA^{\dagger}$$
and the algebra is isomorphic to the Wyel-Heisenberg algebra. Since ${\bf A}_j$'s do not commute, in general, we  have
$$A\mid {\bf A}_{1},{\bf A}_{2},...{\bf A}_{\tau},j\rangle\not={\bf A}_{1}{\bf A}_{2}...{\bf A}_{\tau}\mid {\bf A}_{1},{\bf A}_{2},...{\bf A}_{\tau},j\rangle.$$
Observe that if the matrices ${\bf A}_{1},{\bf A}_{2},...,{\bf A}_{\tau}$ are replaced by complex variables $z_1,...,z_{\tau}$ and the basis is replaced by $\phi_{n_1}\otimes...\otimes\phi_{n_{\tau}}$ then the states (\ref{add2}) can be considered as multi-boson states with no interaction between  bosons. In such a case, when (\ref{mb}) is satisfied, if we define a set of operators similar to the above case we can have
\begin{equation}\label{mb2}
A\mid z_1,...,z_{\tau},j\rangle=z_1...z_{\tau}\mid z_1,...,z_{\tau},j\rangle.
\end{equation}
For the states (\ref{add2}), if we assume that the matrices satisfy $[{\bf A}_j,{\bf A}_k]=0$ for all pairs $j\not=k$ we can have a relation similar to (\ref{mb2}). Such an assumption will be very strong.\\
In order to have an annihilation operator so that the MVCS $\mid {\bf A}_{1},{\bf A}_{2},...{\bf A}_{\tau},j\rangle$ as an eigenvector of it, it has to be defined in such a way that the action of the operator $A$ affects only the first component of the vector, that is,
\begin{equation}\label{mb3}
A \chi^{j}\otimes\phi_{n_1}\otimes...\otimes\phi_{n_{\tau}}=\sqrt{x_{n_1}}\chi^{j}\otimes\phi_{n_1-1}\otimes\phi_{n_2}\otimes...\otimes\phi_{n_{\tau}}.
\end{equation}
In this case the other two operators take the form
\begin{eqnarray}
A^{\dagger} \chi^{j}\otimes\phi_{n_1}\otimes...\otimes\phi_{n_{\tau}}&=&\sqrt{x_{n_1+1}}\chi^{j}\otimes\phi_{n_1+1}\otimes\phi_{n_2}\otimes...\otimes\phi_{n_{\tau}}\label{mb4}\\
N \chi^{j}\otimes\phi_{n_1}\otimes...\otimes\phi_{n_{\tau}}&=&x_{n_1}\chi^{j}\otimes\phi_{n_1}\otimes\phi_{n_2}\otimes...\otimes\phi_{n_{\tau}}.\label{mb5}
\end{eqnarray}
Under the definition (\ref{mb3}) we have
\begin{equation}\label{mb7}
A\mid {\bf A}_{1},{\bf A}_{2},...{\bf A}_{\tau},j\rangle={\bf A}_{1}\mid {\bf A}_{1},{\bf A}_{2},...{\bf A}_{\tau},j\rangle
\end{equation}
and if we assume (\ref{mb}) we again get an algebra isomorphic to the Wyel-Heisenberg algebra.\\
For the states of Section \ref{dep} one can define the operators by the same relations of (\ref{mb3}),(\ref{mb4}),(\ref{mb5}) and get a relation similar to (\ref{mb7}). If we attempt to define operators as (\ref{ooo}) we end at a situation similar to multi-boson case with interaction between bosons.
\section{A physical model}\label{phy}
In the following we adapt the above methods to construct a class
of CS for a Hamiltonian, which is a tensor product of $r$
multi-level quantum Hamiltonians. In the sequel we also
derive CS for tensored Jaynes-Cummings models and for a two-mode
two-level generalization of the Jaynes-Cummings model.

 For each $j=1,2,...,r$, let $H_j$ be a Hamiltonian for an $n$-level atomic system.
 Let $E_{k,m_j}^{j}$ and $\phi_{k,m_j}^{j}~,k=1,...,n;~m_j=1,2,...,\infty$ be the eigenenergies and eigensolutions
 of the Hamiltonian $H_j$ for $j=1,...,r$. Suppose that there is
 no degeneracy and the energies are ordered as follows.
$$0<E_{k,0}^{j}<E_{k,1}^{j}<E_{k,2}^{j}<...;~k=1,2,...,n;~j=1,2,...,r.$$
For each $j=1,2,...,r$ suppose also that the eigensolutions
$\phi_{k,m_j}^{j}~,k=1,...,n;~m_j=1,2,...,\infty$ form an
orthonormal basis of the quantum Hilbert space
$\mathfrak{H}^{j}_{QM}$ of the Hamiltonian $H_j$. Let us rearrange
the energies as follows:
$$\varepsilon_{k,m_j}^{j}=E_{k,m_j}^{j}-E_{k,0}^{j},~~k=1,...,n;~m_j
=0,1,...,\infty;~~j=1,2,...,r.$$
Thereby we have
$$0=\varepsilon_{k,0}^{j}<\varepsilon_{k,1}^{j}<
\varepsilon_{k,2}^{j}<...;~k=1,2,...,n;~j=1,2,...,r.$$
Let $\mathfrak{H_1},...,\mathfrak{H_r}$ be abstract separable
Hilbert spaces with orthonormal bases
$\{\psi_{m_j}^{j}\}_{m_j=0}^{\infty},~j=1,2,...,r$ and
$\{\chi_{k}\}_{k=1}^{n}$ be the natural basis of $\mathbb{C}^{n}$.
Let
$$H=\bigotimes_{j=1}^{r}H_{j},\;\;\;\mathfrak{H}_{QM}=
\bigotimes_{j=1}^{r}\mathfrak{H}_{QM}^{j},\;\;\;\text{and}\;\;\;
\mathfrak{H}=\bigotimes_{j=1}^{r}\mathfrak{H}_j.$$ Then $H$ is a
Hamiltonian describing n level tensored systems with
the state Hilbert space $\mathfrak{H}_{QM}$. The eigenenergies and
the eigensolutions of $H$ can be written as
$$\phi_{k,m_1,...,m_r}=
\bigotimes_{j=1}^{r}\phi_{k,m_j}^{j}\;\;\text{and}\;\;E_{k,m_1,...,m_r}
=\prod_{j=1}^{r}E_{k,m_j}^{j},$$
where $k=1,...,n$. Observe that the energies $E_{k,m_1,...,m_r}$
of the Hamiltonian $H$ can be degenerate. However, it will not
affect the CS construction. Since we sum over all the indices of
the energies and the energies of the individual Hamiltonians
$H_j,~j=1,...,r$ are non-degenerate, the degeneracy of
$E_{k,m_1,...,m_r}$ will not cause any problem in obtaining a
resolution of the identity.
 Define $$V:\mathfrak{H}_{QM}\longrightarrow\mathbb{C}^n\otimes\mathfrak{H}\;\;\text{ by}\;\;\;
 V=\sum_{k=1}^{n}\sum_{m_1=0}^{\infty}\cdots\sum_{m_r=0}^{\infty}\mid\chi_{k}\otimes\psi_{m_1,...,m_r}\rangle\langle\phi_{k,m_1,...,m_r}\mid,$$
where
$$\psi_{m_1,...,m_r}=\bigotimes_{j=1}^{r}\psi_{m_j}^{j}.$$
 Then $V$ is a unitary operator with
$$\phi_{k,m_1,...,m_r}\mapsto\chi_k\otimes\psi_{m_1,...,m_r}$$
for all $k=1,...,n$ and $m_j=0,1,2,....; j=1,...,n$. With the above setup we have
$$VHV^{-1}=H_{D}=\text{diag}(H_{1}^{D},H_{2}^{D},...,H_{n}^{D})$$
a diagonal $n\times n$ matrix, where $H_{k}^{D},~k=1,2,...,n$ are self-adjoint operators on $\mathfrak{H}$ with
$$H_{k}^{D}\psi_{m_1,...,m_r}=E_{k,m_1,...,m_r}\psi_{m_1,...,m_r},\;\;k=1,2,...,n;\;m_j=0,1,2,...$$ Let $z_{k}^{j}\in\mathbb{C}$ and $z_{k}^{j}=r_{k}^{j}e^{\theta_{k}^{j}},~j=1,2,...,r;~k=1,2,...,n$. Let
$$Z_j=\text{diag}(z_1^j,z_2^j,...,z_n^j)\;\;\text{and}\;\;\rho_j(m_j)=\text{diag}(\varepsilon_{1,m_j}^j!,\varepsilon_{2,m_j}^j!,...,\varepsilon_{n,m_j}^j!),$$
where $\varepsilon_{k,m_j}^j!=\varepsilon_{k,1}^j\varepsilon_{k,2}^j\cdots\varepsilon_{k,m_j}^j$. For each $j$ and $k$, suppose that the series
\begin{equation}\label{as1}
\sum_{m_j=0}^{\infty}\frac{(r_{k}^{j})^{2m_j}}{\sqrt{\varepsilon_{k,m_j}^{j}!}}
\end{equation}
is convergent and the radius of convergence is $L_{k}^{j}$. Let
$$D_j=\{(z_1^j,z_2^j,...,z_n^j)\in\mathbb{C}\times\cdots\times\mathbb{C}~\mid~\abs{z_k^j}<L_{k}^{j}\}.$$
Suppose that there are densities $\lambda_{k}^{j}(r_k^j)$ such that
\begin{equation}\label{as2}
\int_{0}^{L_{k}^{j}}(r_k^j)^{2m}\lambda_{k}^{j}(r_k^j)dr_k^j=\varepsilon_{k,m_j}^{j}!;\;\;\;k=1,2,...,n;~j=1,2,...,r.
\end{equation}
Set
$$d\mu(Z_1,...,Z_r)=\frac{1}{(2\pi)^{rn}}\prod_{j=1}^{r}\prod_{k=1}^{n}\lambda_{k}^j(r_k^j)dr_k^jd\theta_k^j$$
on $D_1\times...\times D_r$.
With this setup the set of vectors
\begin{equation}\label{TCS}
\begin{split}
\mid Z_1,Z_2,...,Z_r,k\rangle&=\mathcal{N}(Z_1,...,Z_r)^{-\frac{1}{2}}\sum_{m_1=0}^{\infty}\sum_{m_2=0}^{\infty}\cdots\sum_{m_r=0}^{\infty}\left[\prod_{j=1}^{r}\rho_j(m_j)^{-\frac{1}{2}}Z_j^{m_j}\right]\\
&\hspace{4cm}\cdot\chi_k\otimes\psi_{m_1}^{1}\otimes\psi_{m_2}^{2}\otimes\cdots\otimes\psi_{m_r}^{r}
\end{split}
\end{equation}
forms a set of MVCS, when the states are normalized as
$$\sum_{k=1}^{n}\langle Z_1,Z_2,...,Z_r,k\mid Z_1,Z_2,...,Z_r,k\rangle=1,$$
and the normalization factor is given by
$$\mathcal{N}(Z_1,...,Z_r)=\sum_{m_1=0}^{\infty}\sum_{m_2=0}^{\infty}\cdots\sum_{m_r=0}^{\infty}Tr\left(\prod_{j=1}^{r}\rho_j(m_j)^{-1}\abs{Z_j}^{2m_j}\right).$$
With the argument of Section \ref{gen} and with (\ref{as1}) it can
be seen that $$\mathcal{N}(Z_1,...,Z_r)<\infty.$$
Again with the
argument of Section \ref{gen} and with (\ref{as2}) one can prove
the resolution of the identity in a straightforward way as follows:
\begin{equation}
\begin{split}
&\sum_{k=1}^{n}\int_{D_1\times D_2\times...\times D_r}\mid Z_1,Z_2,...,Z_r,k\rangle\langle Z_1,Z_2,...,Z_r,k\mid\\
&\hspace{2cm}\cdot \mathcal{N}(Z_1,...,Z_r)d\mu(Z_1,...,Z_r)=\mathbb{I}_n\otimes I_{\mathfrak{H}_1}\otimes\cdots\otimes I_{\mathfrak{H}_r}.
\end{split}
\end{equation}

\begin{remark}\label{RR}
Using the technique of \cite{AEG} the diagonal nature of the VCS
(\ref{TCS}) can be removed. For this let $d\nu$ be the normalized
invariant measure of $SU(n)$ and let $U\in SU(n)$. In (\ref{TCS})
let us replace
$$\prod_{j=1}^{r}\rho_j(m_j)^{-\frac{1}{2}}Z_j^{m_j}\;\;\quad
\text{by}\;\;\quad U\prod_{j=1}^{r}\rho_j(m_j)^{-\frac{1}{2}}Z_j^{m_j}U^{\dagger}$$
then a resolution of the identity can be obtained with the measure
$d\mu(Z_1,...,Z_n)d\nu(U)$ and the normalization factor remains
the same. For a detail explanation we refer \cite{AEG}.
\end{remark}
\begin{remark}
Let $\q_1$ and $\q_2$ be two quaternions as in Section \ref{secq}.
Then it is straightforward to see that the eigenvalues of $\q_1$
are $z_1=a_0+ib$ and $\overline{z}_1$ each with multiplicity 2,
where $b=\sqrt{a_1^2+a_2^2+a_3^2}$. Furthermore, there are
orthonormal eigenvectors $\chi_j^{(11)}$ and
$\chi_j^{(12)};~j=1,2$. Let $U_1$ be unitary matrix such that
$\q_1=U_1\text{diag}(z_1,z_1,\overline{z}_1,\overline{z}_1)U_{1}^{\dagger}=U_1Z_1U_1^{\dagger}$.
Let the corresponding counterparts of $\q_2$ be $z_2$ and $U_2$. In
(\ref{TCS}) one could take
$k=4,~r=2,~Z_1=\text{diag}(z_1,z_1,\overline{z}_1,
\overline{z}_1),~Z_2=\text{diag}(z_2,z_2,\overline{z}_2,\overline{z}_2),~\rho_1(m_1)
=\rho(m)\mathbb{I}_4$
and $\rho_2(m_2)=\lambda(l)\mathbb{I}_4$. In particular, if $Z_1$
and $Z_2$ are replaced respectively by
$e^{i\theta}U_1Z_1U_1^{\dagger}$ and
$e^{i\eta}U_2Z_2U_2^{\dagger}$ we have the results of Eq.
\ref{eq20}. The special case $\rho(m)=m!$ and $\lambda(l)=l!$ may
also be of interest.
\end{remark}
For an explicit demonstration, in the following example of the physical
model we construct MVCS for a
Hamiltonian, which is taken as the tensor product of two simple
cases of the Jaynes-Cummings Hamiltonian (JC). As a second
example, we construct VCS for the Hamiltonian (\ref{2J}) with the
spectrum (\ref{2S}) shifted in a suitable manner.
\subsection{Tensored Jaynes-Cummings model}\label{Jay}
Let us construct MVCS for the Hamiltonian $H=H_{w}\otimes H_{c}$,
where $H_{w}$ is the JC in the week coupling limit and $H_{c}$ is the JC
with the exact resonance and zero coupling (as described below).
Information regarding the JC model is extracted from \cite{DH}.
We outline it according to our needs. For a detailed analysis please see \cite{DH}.\\
The JC model describing a spin-$\frac{1}{2}$ system in interaction
with a one-mode magnetic field, in the rotating wave
approximation, can be given by
\begin{equation}
H_{JC}=\omega(a^{+}a^-+\frac{1}{2})\sigma_0+\frac{\omega_0}{2}\sigma_3+\kappa(a^+\sigma_-+a^-\sigma_+)
\end{equation}
where $\omega$ is the field mode frequency, $\omega_0$ the atomic frequency,
$\kappa$ a coupling constant, $a^+$ and $a^-$ are the usual creation
and annihilation operators for the radiation field, $\sigma_{\pm}$
and $\sigma_3$ are associated with the usual
Pauli matrices and $\sigma_0$ is the identity matrix. \\
{\em The Hamiltonian $H_{w}$}:~It is known that the Hamiltonian
$H_{JC}$ can be diagonalized as
$$O^{\dagger}H_{JC}O=H_{D}=\left(\begin{array}{cc}
H_{D(+)}&0\\
0&H_{D(-)}
\end{array}\right),$$
where $O$ is the diagonalizing operator. From the diagonalized form the energy eigenvalues can be obtained as
$$\epsilon_{n}^-=\omega n+\kappa r(n)\;\;\text{and}\;\;\epsilon_{n}^{+}=\omega(n+1)-\kappa r(n+1)$$
where $r(n)=\sqrt{\delta+n}$, $\delta=\left(\frac{\Delta}{2\kappa}\right)^2$ and $\Delta=\omega-\omega_0$, the detuning with $\Delta>0$. Since $\Delta>0$ we have $\epsilon_{n+1}^->\epsilon_n^-$. If $0\leq \kappa/\omega\leq 2\sqrt{\delta+1}$, the energies $\epsilon_n^+$ are strictly increasing and non-degenerate. Let
$$\omega_{\pm}(\kappa)=\frac{\omega\mp\kappa^2}{\Delta}\;\;\text{and}\;\;E_n^{\pm}=\epsilon_n^{\pm}-\epsilon_0^{\pm}.$$
In the weak coupling limit case we expand $E_n^{\pm}$ and by keeping at most terms of order 2 in $\kappa$ we get
$$E_n^{\pm}(\kappa<<<)=\omega_{\pm}(\kappa)n.$$
Let $\psi_{n}^{\pm}$ be the corresponding normalized energy states. In this case we have
$$\rho_{\pm}(n)=E_1^{\pm}E_2^{\pm}...E_n^{\pm}=\left[\omega_{\pm}(\kappa)\right]^n\Gamma(n+1).$$
{\em The Hamiltonian $H_c$}:~In the absence of the oscillating component of the magnetic field ($\kappa=0$) and for the exact resonance ($\Delta=\omega-\omega_0=0$) we take $H_{JC}:=H_c$. In this case it is straightforward to see that
$$\widehat{E}_n^{\pm}=\omega n\;\;\;\text{and}\;\;\;\widehat{\rho}_{\pm}(n)=\omega^n\Gamma(n+1).$$
Let $\phi_n^{\pm}$ be the corresponding normalized energy states.\\
For $j=1,2$, let $z_1^j,z_2^j\in\mathbb{C}$ and $z_k^j=r_k^je^{i\theta_k^j},\;k=1,2;~j=1,2.$ Let
$$Z_1=\text{diag}(z_1^1,z_2^1),~Z_2=\text{diag}(z_1^2,z_2^2),$$
$$\rho_1(m_1)=\text{diag}(\rho_+(m_1),\rho_-(m_1)),\;\;\rho_2(m_2)=\text{diag}(\widehat{\rho}_+(m_2),\widehat{\rho}_-(m_2))$$
and $\rho_1(0)=\rho_2(0)=\mathbb{I}_2.$ Since, for $k=1,2$,
$$\sum_{m_1=0}^{\infty}\frac{(r_k^1)^{2m_1}}{\sqrt{\rho_{\pm}(m_1)}}\;\;\;\text{and}\;\;\;\sum_{m_2=0}^{\infty}\frac{(r_k^2)^{2m_2}}{\sqrt{\widehat{\rho}_{\pm}(m_2)}}$$
converge everywhere we have $L_k^1=L_k^2=\infty.$ From the inverse Mellin transform \cite{E}
$$\int_{0}^{\infty}e^{-ax}x^{s-1}dx=a^{-s}\Gamma(s)$$
and (\ref{as2}) we observe that
$$\lambda_1^1(r_1^1)=2r_1^1\exp{\left(-\frac{(r_1^1)^2}{\omega_+(\kappa)}\right)},\;\;\;\lambda_2^1(r_2^1)=2r_2^1\exp{\left(-\frac{(r_2^1)^2}{\omega_-(\kappa)}\right)}$$
$$\lambda_1^2(r_1^2)=2r_1^2\exp{\left(-\frac{(r_1^2)^2}{\omega}\right)}\;\;\text{and}\;\;\lambda_2^2(r_2^2)=2r_2^2\exp{\left(-\frac{(r_2^2)^2}{\omega}\right)}.$$
In this case a set of MVCS for $H$ can be written as follows:
\begin{equation}\label{2JCS}
\mid Z_1,Z_2,k\rangle=\mathcal{N}(Z_1,Z_2)^{-\frac{1}{2}}\sum_{m_1=0}^{\infty}\sum_{m_2=0}^{\infty}\rho_1(m_1)^{-\frac{1}{2}}Z_1^{m_1}\rho_2(m_2)^{-\frac{1}{2}}Z_2^{m_2}\chi_k\otimes\Phi_{m_1}\otimes\Phi_{m_2};
\end{equation}
$k=1,2,$ where $\{\Phi_{m_1}\}$ and $\{\Phi_{m_2}\}$ are bases for two abstract separable Hilbert spaces $\mathfrak{H}_1$ and $\mathfrak{H}_2$,
$$\mathcal{N}(Z_1,Z_2)=\exp{\left(\frac{(r_1^1)^2}{\omega_+(\kappa)}+\frac{(r_2^1)^2}{\omega_-(\kappa)}+\frac{(r_1^2)^2}{\omega}+\frac{(r_2^2)^2}{\omega}\right)}$$
and a resolution of the identity is satisfied as
\begin{eqnarray*}
&&\sum_{k=1}^{2}\int_{0}^{\infty}\int_{0}^{\infty}\int_{0}^{\infty}\int_{0}^{\infty}\int_{0}^{2\pi}\int_{0}^{2\pi}\int_{0}^{2\pi}\int_{0}^{2\pi}\mid Z_1,Z_2,k\rangle\langle Z_1,Z_2,k\mid d\mu(Z_1,Z_2)\\&&\hspace{7cm}=\mathbb{I}_2\otimes I_{\mathfrak{H}_1}\otimes I_{\mathfrak{H}_2},
\end{eqnarray*}
where
$$d\mu(Z_1,Z_2)=\frac{1}{\omega_+\omega_-\omega^2\pi^4}\prod_{j=1}^2\prod_{k=1}^2r_k^jdr_k^jd\theta_k^j.$$

As a last illustrative example of the construction let us introduce VCS for the Hamiltonian (\ref{2J}) with the spectrum (\ref{2S}).
\subsection{The Hamiltonian (\ref{2J})}
Consider the Hamiltonian $H$ of (\ref{2J}) with $g=0$ and denote it by $H_0$. In this case, the spectrum $E_{\pm}^{n,m}$ of (\ref{2S}) reads
$$E^{n,m}_{\pm}=\omega_1(n+1)+\omega_2(m+1)=e_n^{\pm}+e_m^{\pm}$$
Let $\psi^{\pm}_n\otimes\phi_m^{\pm}$ be the corresponding normalized eigensolutions. Set
$$R(n,m)=\text{diag}(E_{+}^{n,m}!,E_{-}^{n,m}!)=\omega_2^m(\alpha)_m\mathbb{I}_2,$$
where $\alpha=1+[\omega_1(n+1)+\omega_2]/\omega_2$ and $(\alpha)_m=\Gamma(m+\alpha)/\Gamma(\alpha)$ is the Pochhammer symbol. Let
\begin{eqnarray*}
R(n)&=&\text{diag}
\left(e_n^+!,
e_n^-!\right)
=\Gamma(n+2)\omega_1^n\mathbb{I}_2\\
Z&=&\text{diag}\left(z_1,z_2\right)\;\;\text{and}\\
\Z&=&\text{diag}\left(v,\overline{v}\right)
\end{eqnarray*}
where $z_1=re^{i\theta_1},z_2=r_2e^{i\theta_2},v=se^{i\eta},\overline{v}=se^{-i\eta}$ and $\theta_1,\theta_2,\eta\in[0,2\pi)$, where the matrix $\Z$ is chosen in a convenient way (see below).
As before, let $\{\phi_n\}_{n=0}^{\infty}$, $\{\psi_m\}_{m=0}^{\infty}$ be orthonormal bases of two separable abstract Hilbert spaces $\mathfrak{H}_1$ and $\mathfrak{H}_2$ respectively and $\{\chi_1,\chi_2\}$ be the natural basis of $\mathbb{C}^2$. Identify the eigensolutions of $H_0$ to the basis of the abstract Hilbert spaces as follows:
$$\left(\begin{array}{c}\psi_{n}^{+}\otimes\psi_{m}^{+}\\0\end{array}\right):=\chi_1\otimes\phi_n\otimes\psi_m\quad
\left(\begin{array}{c}0\\\psi_{n}^{-}\otimes\psi_{m}^{-}\end{array}\right):=\chi_2\otimes\phi_n\otimes\psi_m.$$
With the above setup we write a set of VCS for the Hamiltonian $H_0$ as follows:
$$\mid Z,\Z,k\rangle=\mathcal{N}(Z,\Z)^{-\frac{1}{2}}\sum_{n=0}^{\infty}R(n)^{-\frac{1}{2}}Z^n\mathcal{N}(\Z,n)^{-\frac{1}{2}}\sum_{m=0}^{\infty}R(n,m)^{-\frac{1}{2}}\Z^n\chi_k\otimes\phi_n\otimes\psi_m,$$
$k=1,2,$ where $\mathcal{N}(\Z,n)$ normalizes the second sum as introduced in \cite{NG} (see Eq. \ref{add1} or \cite{NG,TN}). The normalization requirement
$$\sum_{k=1}^2\langle Z,\Z,k\mid Z,\Z,k\rangle=1$$
yields that
\begin{eqnarray*}
\mathcal{N}(Z,\Z)&=&\sum_{n=0}^{\infty}\left[\left(\frac{r_1^{2n}}{\omega_1^n\Gamma(n+2)}
+\frac{r_2^{2n}}{\omega_1^n\Gamma(n+2)}\right)\mathcal{N}(\Z,n)^{-1}\sum_{m=0}^{\infty}\frac{s^{2m}}{\omega_2^m(\alpha)_m}\right]
\end{eqnarray*}
where
$$\mathcal{N}(\Z,n)=\sum_{m=0}^{\infty}\frac{s_2^{2m}}{\omega_2^m(\alpha)_m}={}_1F_1(1;\alpha;\frac{s^2}{\omega_2})$$
converges everywhere.
Since ${}_1F_1(1;\alpha;\frac{s^2}{\omega_2})\geq 1$ for all $s>0$, we have
\begin{eqnarray*}
\mathcal{N}(Z,\Z)&=&\sum_{n=0}^{\infty}\frac{1}{{}_1F_1(1;\alpha;\frac{s^2}{\omega_2})}\left[\left(\frac{r_1^{2n}}{\omega_1^n\Gamma(n+2)}
+\frac{r_2^{2n}}{\omega_1^n\Gamma(n+2)}\right)\right]\\
&\leq&\sum_{n=0}^{\infty}\left[\left(\frac{r_1^{2n}}{\omega_1^n\Gamma(n+2)}
+\frac{r_2^{2n}}{\omega_1^n\Gamma(n+2)}\right)\right]\\
&=&\frac{\omega_1}{r_1^2}(e^{r_1^2/\omega_1})+\frac{\omega_1}{r_2^2}(e^{r_2^2/\omega_1}),
\end{eqnarray*}
which converges everywhere. Let
$$\lambda(r)=\frac{2r^2}{\omega_1^2}e^{-r^2/\omega_1^2},\;\;\;\widehat{\lambda}(s,n)=\frac{2s^{2\alpha-1}}{\omega_2^{\alpha}\Gamma(\alpha)}e^{-s^2/\omega_2}.$$
Then we have
\begin{eqnarray*}
&&\int_0^{\infty}r^{2n}\lambda(r)dr=\omega_1^n\Gamma(n+2),\\
&&\int_0^{\infty}s^{2m}\widehat{\lambda}(s,n)ds=\omega_2^m(\alpha)_m
\end{eqnarray*}
or
$$\int_0^{\infty}r^{2n}\lambda(r)\int_0^{\infty}s^{2m}\widehat{\lambda}(s,n)dsdr=\omega_1^n\Gamma(n+2)\omega_2^m(\alpha)_m.$$
Let $D=[0,\infty)\times[0,2\pi)$ then by setting $n=0$ and $m=0$ in the above integrals one can see that
$$\frac{1}{(2\pi)^4}\int_{D\times D\times D\times D}\lambda(r_1)\lambda(r_2)\widehat{\lambda}(s,n)^2 dr_1d\theta_1dr_2d\theta_2dsd\eta dsd\eta=1.$$
Thereby we have a resolution of the identity
$$\sum_{k=1}^{2}\int_{D\times D\times D\times D}\mid Z,\Z,k\rangle\langle Z,\Z,k\mid d\mu(Z,\Z)=\mathbb{I}_2\otimes I_{\mathfrak{H}_1}\otimes I_{\mathfrak{H}_2},$$
where
$$d\mu(Z,\Z)=\frac{1}{(2\pi)^4}\prod_{j=1}^{2}\mathcal{N}(Z,\Z)\lambda(r_j)\mathcal{N}(\Z,n)\widehat{\lambda}(s,n)^2dr_jd\theta_jdsdsd\eta d\eta.$$
Here again Remark \ref{RR} is applicable. For the Hamiltonian $H_0$ we can have another set of CS similar to the CS (\ref{2JCS}) by appropriately shifting the spectrum. We describe it in the following remark.
\begin{remark}
 Let us shift the spectrum as follows:
\begin{eqnarray*}
e_+^{n,m}&=&E_+^{n,m}-E_+^{n,0}=e_-^{n,m}=E_-^{n,m}-E_-^{n,0}=\omega_2m\\
\epsilon_+^{n,m}&=&E_+^{n,m}-E_+^{0,n}=\epsilon_-^{n,m}=E_-^{n,m}-E_-^{0,n}=\omega_1n.
\end{eqnarray*}
Let
\begin{eqnarray*}
R(n,m)&=&\text{diag}\left(
e_{+}^{n,m}!,e_{-}^{n,m}!\right)
=m!\omega_2^m\mathbb{I}_2\\
R(n)&=&\text{diag}
\left(\epsilon_{+}^{n}!,
\epsilon_{-}^{n}!\right)
=n!\omega_1^n\mathbb{I}_2\\
Z&=&\text{diag}\left(z_1,z_2\right)\;\;\text{and}\\
\Z&=&\text{diag}\left(v_1,v_2\right)
\end{eqnarray*}
where $z_1=re^{i\theta_1},z_2=r_2e^{i\theta_2},v_1=s_1e^{i\eta_1},v_2=s_2e^{i\eta_2}$ and $\theta_1,\theta_2,\eta_1,\eta_2\in[0,2\pi)$.
With the above setup we write a set of VCS for the Hamiltonian $H_0$ as follows:
$$\mid Z,\Z,k\rangle=\mathcal{N}(Z,\Z)^{-\frac{1}{2}}\sum_{n=0}^{\infty}R(n)^{-\frac{1}{2}}Z^n\sum_{m=0}^{\infty}R(n,m)^{-\frac{1}{2}}\Z^n\chi_k\otimes\phi_n\otimes\psi_m,~~k=1,2.$$
By following the details of the Subsection \ref{Jay} we can obtain the normalization factor and the resolution of the identity.
\end{remark}
\section{Conclusion}
The results of \cite{KA} have been generalized with multiple matrices. The results presented in this article serve as a generalization of \cite{NG} since they can be employed for the purpose of obtaining CS for tensored solvable matrix Hamiltonians as well as to obtain CS for multi-level multi-mode solvable systems. Further the results of Section \ref{phy} may also be considered as a generalization of the physical model presented in \cite{AEG}. In order to demonstrate the physical model, in Example \ref{Jay} we have considered the tensor product of two simple cases of the Jaynes-Cummings model. The model can also be applied to other tensored systems, where to solve the moment problem one may adapt the results or the methods of \cite{GPS}. Furthermore, the methods presented here may facilitate the derivation of CS for super Hamiltonians with multiple degrees of freedom. The complex and real representation of quaternions and the real representation of octonions are used in the construction of multi-matrix VCS in several ways, namely the variables $J_{j}$s of (\ref{add1}) are replaced by matrices, the $\rho_j(n_j)$s are replaced by matrices, within the multiple sums one sum is allowed to depend on the other. In the same spirit, with the aid of \cite{AEG}, it is possible to extend the construction to the matrix representation of any Clifford algebra and to any Cartesian product of classical matrix domains.
\section*{Acknowledgements}
Partial financial support of this work from the Natural Sciences and Engineering Research Council of Canada is gratefully acknowledged (A.K., K.T.).


\begin{thebibliography}{XXXX}

\bibitem{AAG} Ali S.T., Antoine J-P., Gazeau J-P., {\em Coherent States, Wavelets and Their Generalizations}, Springer, New York (2000).
\bibitem{GS} Klauder J.R, Skagerstam B.S {\em Coherent States, Applications in Physics and Mathematical Physics},
World Scientific, Singapore, (1985).
\bibitem{Pr} Perelomov A.M.,
{\em Generalized Coherent States and Their Applications}, Springer-Verlag, Berlin, (1986).
\bibitem{Gk} Gazeau J-P., Klauder J.R., {\em J. Phys. A:  Math. Gen.} {\bf 32} 123-132 (1999).
\bibitem{NG}Novaes M., Gazeau J-P.,
{\em J. Phys. A: Math. Gen.} {\bf 36} 199-212 (2003).
\bibitem{KA}Thirulogasanthar K., Ali S.T., {\em J. Math. Phys.} {\bf 44}, 5070-5083 (2003).
\bibitem{AEG} Ali S.T., Englis M., Gazeau J-P., {\em J. Phys. A: Math. Gen.} {\bf 37} 6067-6089 (2004).
\bibitem{AGM} Antoine J-P., Gazeau J-P., Monceau P., Klauder J.R., Penson K.A., {\em J. Math. Phys.} {\bf 42}, 2349-2387 (2001).
\bibitem{F} Fakhri H., {\em Phys. Lett. A} {\bf 313} 243-251 (2003).
\bibitem{TN} Thirulogasanthar K., Saad N., von Keviczky A.B., {\em J. Math. Phys.} {\bf 45} 2694-2717 (2004).
\bibitem{K} Klauder J.R., {\em J. Phys. A: Math. Gen.} {\bf 29} L293-L298 (1996).
\bibitem{DH} Daoud M., Hussin V., {\em J. Phys. A: Math. Gen.} {\bf 35} 7381-7402 (2002).
\bibitem{JA} Janowicz M.W., Ashbourn J.M.A., {\em Phys. Rev. A.} {\bf 55} 2348-2359 (1997).
\bibitem{As} Ashraf M.M., {\em Phys. Rev. A} {\bf 50}  5116-5121 (1994).
\bibitem{Gao} Gao Y.F., Feng J., Shi S.R., {\em Internat. J. Theoret. Phys.} {\bf 41} 867-875 (2002).
\bibitem{GPS} Klauder J.R, Penson K., Sixdeniers J-M., {\em Phys. Rev. A.} {\bf 64} 013817 (2001).
\bibitem{key6}
Yongge Tian, {\em Adv. Appl. Clifford Algebras} {\bf 10} 61-90 (2000).
\bibitem{E} Erdelyi A., Magnus W., Oberhettinger F. and Tricomi F.G., {\em Table of integral transforms,} New York, McGraw-Hill (1953).

\end{thebibliography}
                      \end{document}